\documentclass[letterpaper,twocolumn,10pt]{article}
\usepackage{usenix2022}

\usepackage{tikz}
\usepackage{amsmath}
\usepackage{amssymb}
\usepackage{algorithm}
\usepackage{algorithmic}
\usepackage{multirow}
\usepackage{makecell}
\usepackage{bbding}
\usepackage{xcolor}

\newcommand{\louis}[1]{\textcolor{black}{#1}}
\newcommand{\lyf}[1]{\textcolor{black}{#1}}
\newcommand{\xqf}[1]{\textcolor{black}{#1}}

\usepackage{filecontents}
\usepackage{booktabs}
\usepackage{tikz}

\usepackage{enumitem}
\setlist[itemize]{leftmargin=*}
\setlist[enumerate]{leftmargin=*}

\begin{document}
%-------------------------------------------------------------------------------

%don't want date printed
\date{}

% \title{Exorcising ``Wraith'': Defending LiDAR-based Object Detector 
% in \\ Automated Driving System against Appearing Attacks}

\title{Exorcising ``Wraith'': Protecting LiDAR-based Object Detector \\ 
in Automated Driving System from Appearing Attacks}

%for single author (just remove % characters)
% \author{\IEEEauthorblockN{Anonymous Author(s)}}
\author{\rm{Qifan Xiao*, Xudong Pan*, Yifan Lu, Mi Zhang\textsuperscript{\Envelope}, Jiarun Dai, Min Yang\textsuperscript{\Envelope}} \\ \textit{Fudan University, China} \\ 
\{20210240056@, xdpan18@, luyifan21@m., jrdai14@, mi\_zhang@, m\_yang@\}fudan.edu.cn \\ 
{\small{(*: co-first authors; \Envelope: corresponding authors)}}
}

% \author{
% {\rm Qifan Xiao}\\
% Fudan University
% \and
% {\rm Xudong Pan}\\
% Fudan University
% \and
% {\rm Yifan Lu}\\
% Fudan University
% \and
% {\rm Mi Zhang}\\
% Fudan University
% \and
% {\rm Jiarun Dai}\\
% Fudan University
% \and
% {\rm Min Yang}\\
% Fudan University
% }

\maketitle

\begin{abstract}
Automated driving systems rely on 3D object detectors to recognize possible obstacles from LiDAR point clouds. However, recent works show the adversary can forge non-existent cars in the prediction results with a few fake points (i.e., \textit{appearing attack}). By removing statistical outliers, existing defenses are however designed for specific attacks or biased by predefined heuristic rules. Towards more comprehensive mitigation, we first systematically inspect the mechanism of 
% previous
\xqf{recent}
appearing attacks: Their common weaknesses are observed in crafting fake obstacles which (i) have obvious differences in the local parts compared with real obstacles and (ii) violate the physical relation between depth and point density. 

In this paper, we propose a novel plug-and-play defensive module which works by side of a trained LiDAR-based object detector to eliminate forged obstacles where a major proportion of local parts have low \textit{objectness}, i.e., to what degree it belongs to a real object. At the core of our module is a \textit{local objectness predictor}, which explicitly incorporates the depth information to model the relation between depth and point density, and predicts each local part of an obstacle with an \textit{objectness} score. Extensive experiments show, our proposed defense eliminates at least $70\%$ cars forged by three known appearing attacks in most cases, while, for the best previous defense, less than $30\%$ forged cars are eliminated. Meanwhile, under the same circumstance, our defense incurs less overhead for AP/precision on cars compared with existing defenses. Furthermore, We validate the effectiveness of our proposed defense 
% on real-world PC data obtained from the D-KIT Advanced with Velodyne-128 in a closed road environment and 
on simulation-based closed-loop control driving tests in the open-source system of Baidu's Apollo.
\end{abstract}
%to handle the complexity of real-world driving scenes. 
\section{Introduction}\label{sec:Introduction}
% In the past decade, deep learning empowers ma.ny real-world application in, e.g., finance \cite{Heaton2016DeepLF}, healthcare \cite{Esteva2017DermatologistlevelCO} and transportation \cite{edward2016doctorai,jiaxuan2019anAutomated}, 
In automated driving systems (ADS), multiple deep neural networks (DNNs) are jointly deployed to provide key functionalities of localization, perception and planning, stimulating the recent development of automated transportation \cite{marco2016MLinTrack,sina2020practical,yunli2021MLEnabled}. The robustness of each DNN module is of key importance to the security of the whole ADS. A typical example is the \textit{perception} module, which relies on a vector of \textit{object detectors}, based on multiple sources like cameras and LiDARs \cite{LiDAR}, to predict the categories and locations of the obstacles around the ADS \cite{yulong2019advObj, gregory2019lasernet}. 
As the LiDAR point clouds (PCs) contain richer location information than the images from cameras, most commercial ADS, including Google's Waymo One \cite{Waymo, waymo_report} and Baidu's Apollo \cite{Apollo, apollo_code}, set LiDARs as the main sensors and rely on the detection results of \textit{LiDAR-based object detectors} for obstacle perception  \cite{alex2019pointpillars,shaoshuai2019pointrcnn,shaoshuai2020pvrcnn,gregory2019lasernet}.

%%%%%%% BEGIN ATTACK SCENARIO
\begin{figure}[t]
    \centering
    \includegraphics[width=0.45\textwidth]{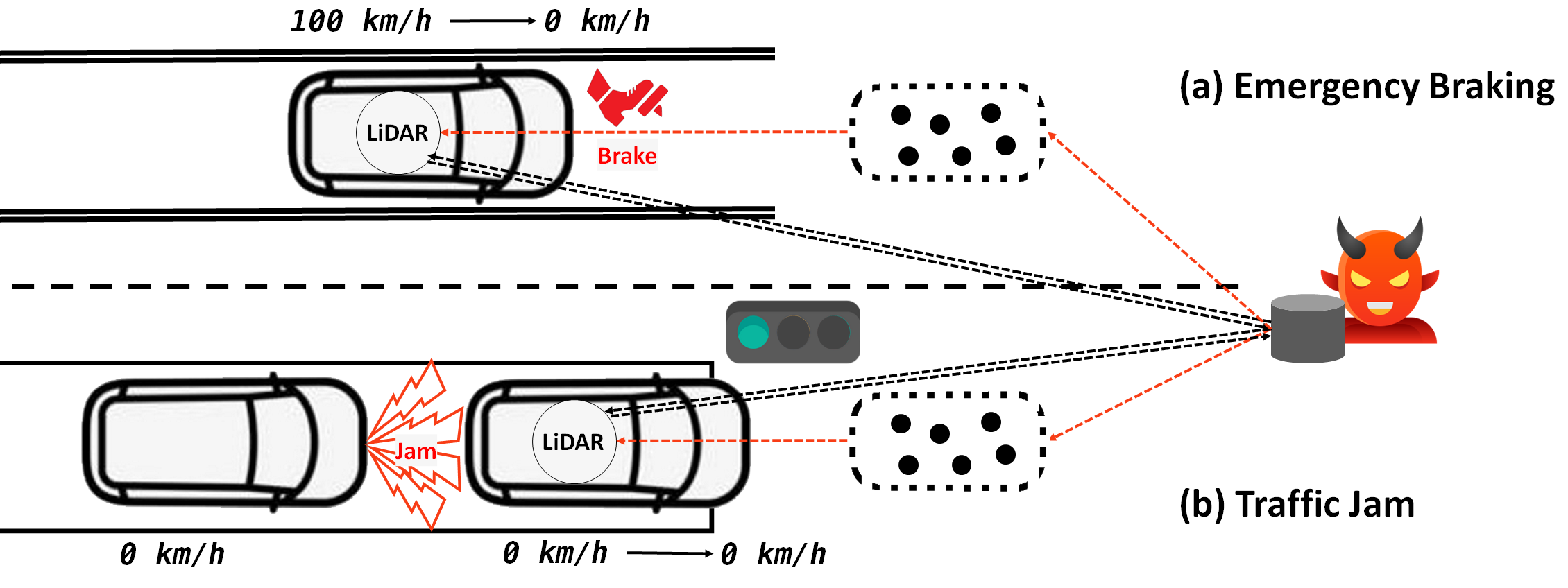}
    \caption{Appearing attacks on LiDAR-based object detectors in ADS can cause severe traffic accidents by forging cars.}
    \label{fig:atk_scenario}
\end{figure}
%%%%%%%% END ATTACK SCENARIO

Differently using PC as the model input, LiDAR-based object detectors still share the common vulnerability against \textit{adversarial examples} \cite{Szegedy2014IntriguingPO,yulong2019advObj,kaichen2021robust}. In general, the attacker can spoof the LiDAR sensors with a limited number of perturbed/crafted  points to mislead the detector's prediction.
As a popular attack class, the \textit{appearing attack} aims at forging non-existent cars in the detection results to cause traffic jams and emergency braking \cite{hocheol2017illusion,yulong2019advSensor} (Fig.\ref{fig:atk_scenario}).
Despite the severity, existing defenses \cite{jiachen2020towards,qi2020object, zhongyuan2021shadowcatcher} either have strong prior assumptions on the undergoing attacks, or are biased by predefined heuristic rules, insufficient for handling complex driving scenarios (\S\ref{sec:Limitation}).

% %%%%%%%%%% BEGIN GENERAL DEFENSE
% \begin{figure}[ht]
%     \centering
%     \includegraphics[width=0.45\textwidth]{figs/general_defense_2.png}
%     \caption{The mechanism of our plug-and-play defense.}
%     \label{fig:general_defense}
% \end{figure}
% %%%%%%%%%%%% END GENERAL DEFENSE

\noindent\textbf{Our Work.} In this paper, we propose a novel plug-and-play defense for 3D object detectors, which, instead of constructing a more robust model,
adopts a \textit{local objectness predictor} (LOP) module to detect and eliminate forged obstacles from the original detection results. 
% (Fig.\ref{fig:general_defense}).
In general, our LOP is designed as a point-wise PC classifier \cite{charles2017PN++,yangyan2018PCNN,charles2017PN,yue2019DGCNN} which learns to predict each local part of a detected object with an \textit{objectness} score, i.e., the confidence of whether the local part belongs to a real object.
By systematizing \xqf{recent} appearing attacks, we develop the following defensive insights: 
\begin{enumerate}[leftmargin=*]
    \item \xqf{Recent} appearing attacks focus on increasing the confidence score of a fake detection result without considering the local difference between a real and a forged obstacle. Although an increased confidence score enhances the possibility of a non-existent obstacle to be detected by a 3D object detector, most appearing attacks leave the fake and the real obstacles locally distinguishable when inspected at the granularity of pillars or voxels (\S\ref{sec:method:preparation}) 
    \item Constrained by the physical capability of attack apparatus, appearing attacks are usually unable to forge a fake obstacle without violating some physical laws,
    especially the inimitable relation between the depth and the point density of real obstacles \cite{depth_density}.
    To pose real-world threats, the forged obstacles have to be close to the victim ADS, because otherwise they can be easily bypassed after the victim's re-routing.
    % \louis{In order to realize the attack, the forged obstacles have to be close to the victim ADS because otherwise they could be easily bypassed after the victim's re-routing.}
    % In terms of the attacker's goal, the forged obstacles have to be located near the victim ADS. Otherwise, the victim ADS can easily re-route and bypass the fake obstacles.
    Yet, constrained by the attack apparatus (e.g., a laser transmitter \cite{jiachen2020towards}), the attacker can only forge a limited number of points near the victim during one scan of the LiDAR, which could hardly reach the normal point density of a real car at a close distance (\S\ref{sec:method:lop}). 
    % \xqf{Besides, though there exists a defense method \cite{zhongyuan2021shadowcatcher} which also discovered a similar physical law between the depth and the point density of real obstacles, they decided to turn it into a heuristic design based on their observation, and we further prove that it is not robust enough compared with LOP in the later experiments (\S\ref{sec:Experiment:results:comp_attack}).}

\end{enumerate}

Concurrent to our defense, Hau et al. \cite{zhongyuan2021shadowcatcher} also notices the importance of the physical law in detecting forged obstacles, and presents a set of hand-crafted rules to eliminate the anomaly. Our work steps further by showing stronger robustness can be achieved if we exploit learning-based techniques to model the complicated physical laws. In fact, modeling the relation between the depth and the point density is rather challenging with hand-crafted rules. For example, although most of the real cars with smaller depth tend to have larger point density, those real cars occluded by others may also have smaller depth and point density simultaneously (Fig.\ref{fig:depth_density}). 
To address this challenge, we implement the LOP as a DNN-based point-wise PC classification model and explicitly incorporate the depth information of each point into its feature vector.
This substantially improves the modeling capability compared with using the original input feature for statistical outlier detection.

Moreover, another technical challenge is the lack of no explicit annotation available for supervising the training of LOP in standard 3D object detection datasets. 
Inspired by a recent observation that a single part of the input already contains rich semantic information for a PC model to predict its related object's category and location \cite{qi2020object},
we construct a self-supervised learning task where the LOP learns to predict whether a pillar intersects with any bounding box of real objects based on the features of its inside points.
During the detection, we first divide the input 3D space into equal-sized pillars, then the LOP predicts an objectness score for each pillar intersected with a predicted object's bounding box. By majority voting on the local objectness predictions, our defense determines whether the object is real or fake (\S\ref{sec:method:voting}).

% For experiments, we extensively evaluate our proposed defense against three known appearing attacks on three mainstream LiDAR-based object detectors, i.e., PointPillars \cite{alex2019pointpillars}, PointRCNN \cite{shaoshuai2019pointrcnn} and PV-RCNN \cite{shaoshuai2020pvrcnn} on the KITTI dataset \cite{andreas2012kitti} and on real-world PC data we collect from a driving test of the D-KIT Advanced with Velodyne-128 \cite{DKit} in a closed road environment. Moreover, we empirically validate that the effectiveness of our proposed LOP is robust to both the architecture design of the LOP and the type of the LiDAR-based object detector under guard, which further implies our defense is more general-purpose than existing defenses (\S\ref{sec:Experiment}).

% % with and without our LOP under different
% \louis{attacks}.
% Our results show that with the deployment of our LOP, \louis{at least} $70\%$ of the fake cars forged by existing appearing attacks are eliminated,
% while there are only less than $30\%$ forged cars are eliminated for the best previous defense in certain cases

% while the precision on cars is only degraded by less than $3\%$ in most cases,  which is far more less than other defenses. Furthermore, we also validate our defense on real-world PC data obtained from a D-KIT Advanced car with Velodyne-128 in a closed road environment.

% data0->不防御的情况下，受攻击的3D实体检测模型的precision范围
% data1->防御的情况下，受攻击的3D实体检测模型的precision范围
% data2->不防御的情况下，3D实体检测模型的正常precision范围
% data3->防御的情况下，3D实体检测模型的正常precision范围

\noindent\textbf{Our Contributions.} In summary, the key contributions of this work are as follows:

\noindent\textbullet\ 
% We for the first time inspect and systematize the common limitations of existing appearing attacks, 
We systematize the limitations of 
% existing 
\xqf{recent} appearing attacks in violating the physical invariants and propose a learning-based defense to detect the forged obstacles with anomaly in the relation between the depth and the point density for the mainstream LiDAR-based object detectors.
% which inspire us to derive a general defense method for a generic set of mainstream LiDAR-based object detectors under multiple attack techniques.

\noindent\textbullet\ We propose the design of our local objectness predictor (LOP) which learns to predict the confidence of whether a local object part belongs to a real object, and allows plug-and-play integration with different defense targets for enhancing robustness against popular appearing attacks. 

% Our LOP can be constructed with no external annotation and achieves effectiveness independent of the LOP architecture and the type of the LiDAR-based object detector under guard.

\noindent\textbullet\  Extensive evaluation on mainstream 3D detectors (i.e., PointPillars \cite{alex2019pointpillars}, PointRCNN \cite{shaoshuai2019pointrcnn} and PV-RCNN \cite{shaoshuai2020pvrcnn}) on the KITTI dataset \cite{andreas2012kitti} and on real-world PC data we collect from a driving test of the D-KIT Advanced with Velodyne-128 \cite{DKit} validate the advantages of our proposed defense under three popular attacks. For example, with the same-level trade-off in model utility, our proposed defense eliminates at least $70\%$ cars forged by most appearing attacks, while the best baseline method only eliminates the forged ones less than $30\%$.

\noindent\textbullet\ Moreover, we empirically validate that the effectiveness of our proposed LOP is robust to the architecture design of the LOP, the type of the defense target (including fusion models) which further implies our defense is more general-purpose than existing defenses. Besides, we also provide a preliminary study on the robustness of LOP against adaptive attacks. 

\noindent\textbullet\ 
% Meanwhile, we also validate our proposed defense on PC data gathered with a real car driven by ADS 
We further implement and evaluate the effectiveness of LOP in Apollo 6.0.0, an end-to-end open-source self-driving system, with closed-loop control in the LGSVL simulation tests, which validates the system-level usefulness of our proposed defense in both benign and adversarial scenarios.

% Meanwhile, under the same circumstances, our defense brings uniformly less decrease in AP/precision on cars than other defenses.
% {For example, our LOP increases a 3D object detectors' precision under normal circumstances or under attack.}
% \input{tex/relatedwork}
\section{Background}\label{sec:Background}
\noindent\textbf{Basics of LiDAR.} As one of the main sensors deployed in an automated driving system (ADS), a LiDAR (Light Detection and Ranging) scans the surrounding environment and generates a point cloud (PC) $X=\{(x_i,y_i,z_i,int_i)\}\in R^{n\times 4}$, including $n$ points with $(x_i,y_i,z_i)$ as $i$-th point's location and $int_i$ as $i$-th point's intensity, during each detection \cite{LiDAR, hocheol2017illusion}. 
Technically, the LiDAR first emits a laser ray consecutive in both horizontal and vertical directions, which then captures the reflected lasers, records their time of flight and light intensity, and further computes the depth and 3D coordinate of the points related to these reflected lasers. 
Finally, the LiDAR collects these information to generate the raw PC, which represents the object surfaces in the surrounding environment, and sends this raw PC to the ADS for downstream processing. 

% Because the time of flight and light intensity of reflected lasers are hardly affected by the change of object surface texture, the traditional image-based adversarial attack methods have limited effects on these LiDAR-based 3D object detectors. On the other words, LiDARs are more robust to the change of environment and the image-based adversarial attacks than cameras. This is primarily why most commercial self-driving systems, including Google's Waymo One \cite{Waymo} and Baidu's Apollo \cite{Apollo}, mainly rely on the sensor data (e.g., PC) generated from LiDAR for obstacle perception.

\noindent\textbf{3D Object Detectors.}
% Options: Draw a picture to introduce the meanings of bounding box, pillar and voxel
DNN-based 3D object detectors empower modern ADS for perceiving and detecting objects in the surrounding environment (i.e., \textit{obstacle perception}).
Technically, a 3D object detector usually takes PC as the input and returns the category and \textit{bounding box}, a rectangle or cuboid which bounds the detected object to represent its location in a PC, of each perceived object \cite{yulan2021deep}. 
In most cases, 3D object detectors can be regarded as the combination of three modules: the preprocessing, the backbone and the prediction modules. 

A typical preprocessing module first divides the points of PC into a number of sets (e.g., voxels or pillars) based on specific rules and then calculates the statistical information \cite{alex2019pointpillars,yin2018VoxelNet}, or uses DNN models, such as PointNet \cite{charles2017PN} or DGCNN \cite{yue2019DGCNN}, to generate the feature vectors for each point \cite{shaoshuai2019pointrcnn}. Then,
the backbone module implemented with 2D/3D convolutional neural networks (CNN) \cite{yann1989backpropagation, shuiwang20103DCNN} extracts the PC's features and generates the global feature map. Finally, the prediction module in one-stage 3D object detectors like VoxelNet \cite{yin2018VoxelNet} and PointPillars \cite{alex2019pointpillars} directly predicts the bounding box and category of each obstacle based on the global feature map. Differently, in two-stage 3D object detectors like PointRCNN \cite{shaoshuai2019pointrcnn} and PV-RCNN \cite{shaoshuai2020pvrcnn}, the prediction module predicts the proposal bounding boxes of objects based on the global feature map and generate a local feature map for each object based on the combination of the global feature map and the related proposal bounding boxes at the first stage, and then the final bounding box and category of each obstacle based on each local feature map at the second stage.

% Considering the direct effect on the future
% travel plan,
% % \louis{driving plan, }
% the location prediction % from the detection results of a LiDAR-based object detector
% of the detection results
% is much more important than the predicted category at the self-driving level. The key to predicting the bounding box is the spatial information of the environment, which is stored explicitly in PC while implicitly in images. In other words, a PC always contains more spatial information than an image, which explains why the bounding box prediction from a LiDAR-based object detector is usually considered more reliable than that from an image-based object detector \cite{joseph2016yolo, shaoqing2015fasterrcnn, wei2016ssd}.

\noindent\textbf{Adversarial Example.} In general, given a machine learning model $F$ and a normal sample $x$ with label $y$, an adversarial example $x'$ is generated from $x$ by adding a slight perturbation to mislead the victim model's prediction while causing no modification to either the model's architecture or the parameters \cite{Szegedy2014IntriguingPO,abdullah2020advpc,yue2020onisomety}.
According to the attack goal, an adversarial example can be further categorized into untargeted and targeted. By definition,
an untargeted attack aims at misleading the victim model into $F(x')\neq y$, while a targeted attack aims at misleading the victim model into $F(x')=y'$, where $y'$ is the target label specified by attacker. 
According to \cite{nicholas2017towards}, the targeted adversarial attack can be further represented as the optimization problem:
% \begin{align}
% & {argmin}_{x'}||x-x'||_p \label{equal:CW} \\ 
% \text{s.t.}\quad & F(x')=y'\text{ and } x' \in X \notag
% \end{align}
\vspace{-1mm}
\begin{align}
& {argmin}_{x'}||x-x'||_p & \text{s.t.} \quad F(x')=y'\text{ and } x' \in X \label{equal:CW}
\end{align}
\vspace{-1mm}
where the objective $\min\|x - x'\|$ restricts the region of perturbation (i.e., attack budget) and $X$ denotes the input space. In the context of ADS, to cause severe safety issues, several recent adversarial attacks focus on conducting LiDAR spoofing to forge a non-existent object in the detection results of a LiDAR-based object detector, or called \textit{appearing attacks}, on which we provide a detailed review in Section \ref{sec:attack_review}.

% A very recent line of works showed that PC models, including those 3D object detectors are vulnerable to \louis{adversarial attacks}. Xiang et al. suggested the first adversarial attack that \louis{against} PC models \cite{chong2019generating}. Tsai et al. and Hamdi et al. both focused on the transferability of adversarial samples, and suggested two different attack methods that the adversarial samples generated by them can affect the prediction of multiple PC classifiers at the same time \cite{abdullah2020advpc,yue2020onisomety}.
\section{Security Settings}\label{sec:ThreatModel}
% In this section, we first define the security settings of an appearing attack. Then, we review the existing attack and defense techniques. Finally, we further discuss the limitations of existing defenses in countering existing appearing attacks.

\subsection{Threat Model}
\noindent$\bullet$\textbf{ Attacker's Goal}. In general, the direct goal of an appearing attack is to forge fake cars, in the detection results of the LiDAR-based object detector in ADS. To refine the attack goal above, we first analyze the following two attack scenarios of an appearing attack. 

\noindent\textbf{Attack Scenario 1.} \textit{(On the Highways)}  As shown in the top part (a) of Fig.\ref{fig:atk_scenario}, an attacker can spoof the LiDAR of the victim ADS when it passes by. Detecting a forged car at the immediate front, the victim will make a stop decision and decrease its speed to 0 km/h within seconds. The unpredictable emergency braking may leave no reaction time for other vehicles behind. This may lead to a rear-end collision or even more severe traffic accidents.

\noindent\textbf{Attack Scenario 2}. \textit{(At the Traffic Lights)} Similarly, as shown in the bottom part (b) of Fig.\ref{fig:atk_scenario}, the attacker conducts LiDAR spoofing when the victim ADS stops at the red light. By forging a fake car ahead, the victim will keep immobile even after the traffic signal turns green, blocking other vehicles behind and causing a traffic jam.

As the two attack scenarios show, to cause a real-world threat, the forged cars are required to be not only recognized by LiDAR-based object detectors with sufficiently large confidence scores, but also close enough to result in the re-routing of the victim. Therefore, we further refine the attack goal to expect the cars to be forged in a close distance to the victim. Specifically, in this work we require a forged car to be within a $5\sim{10}$ meters to the victim to pose a sufficient threat \cite{yulong2019advSensor, jiachen2020towards}.

\noindent$\bullet$\textbf{ Attacker's Capability.}
Following the threat model in recent attacks \cite{yulong2019advSensor, jiachen2020towards}, our defense mainly aims at mitigating an attacker satisfying the following threat model: 

\noindent \textbf{Assumption 1.} \textit{(Prior Knowledge)} The attacker knows the architecture and the parameters of the LiDAR-based object detector deployed on the victim ADS (i.e.,  \textit{white-box}).

% It is feasible because most ADS are open-sourced on third-party platforms (e.g., GitHub), where the attacker can audit the source code to figure out the details of the target model. 

\noindent \textbf{Assumption 2.} \textit{(Number of Added Points)} The attacker can inject at most 200 points (according to \cite{jiachen2020towards}) into the input PC of the victim 3D object detector in one scan of LiDAR.

% , the 200 points are the upper bound on the capability of existing physical equipment for LiDAR spoofing in one scan.

\noindent \textbf{Assumption 3.} \textit{(Features of Added Points)} The attacker is allowed to inject points at any location and with arbitrary light intensity, which is imposed for a more generic defense.

% . Although there exists a bounded feasible region for LiDAR spoofing due to the capability of physical equipment, we relax this requirement

\noindent$\bullet$\textbf{ Attack Process.} Before the attack starts, the attacker deploys a physical equipment to receive the lasers emitted by the victim ADS's LiDAR, and shoot lasers back to the LiDAR. 
% The details of this equipment are described in Appendix \ref{appendix:physical_equipment}. 
% With the support of this equipment, the attacker can inject a certain number of points within a certain area into the LiDAR's generated PC as he/she wishes. 
Later, the LiDAR-based 3D object detectors of the victim takes the infected PC and predicts a non-existent car. 
Finally, the victim re-routes to avoid the non-existent car, which may lead to severe collision accidents.

\subsection{Recent Appearing Attacks}
\label{sec:attack_review}
Next, we review the recent appearing attacks on LiDAR-based object detectors. As one of the earliest work, Shin et al. propose a spoofing attack by randomly injecting points into a certain area regardless of the LiDAR-based object detectors of the victim ADS, which is sufficient to forge a non-existent car \cite{hocheol2017illusion}. Inspired by Shin's work, Cao et al. standardize the attack pipeline of adversarial spoofing attack, and propose an appearing attack, Adv-LiDAR, which aims at breaking Apollo's detection system \cite{yulong2019advSensor}. By modeling the preprocessing and postprocessing modules in Apollo's LiDAR-based object detector, Adv-LiDAR successfully uses traditional adversarial attack technology to forge non-existent cars. However, Sun et al. later prove that other 3D object detectors such as PointPillars and PointRCNN will not be affected by the the adversarial samples generated by Adv-LiDAR, and then suggest a more general black-box appearing attack based on the intrinsic physical nature of LiDARs \cite{jiachen2020towards}. Also, another attack by Yang et al. shares the same attack goal as the above appearing attacks but uses a different attack process and physical equipment \cite{kaichen2021robust}. Specifically, they use a physical object which is specially designed to tempt the 3D object detector to predict itself as a car with a falsely enlarged bounding box and therefore fabricate a non-existent part of this object in the model's perception. For completeness, we also cover this attack into the appearing attacks in experiments. 

% Despite the different attack strategies, we observe that most of them rely solely on increasing the confidence score of the fake cars in the detection results for car forgery, which may leave a clear differences between local parts of a real and a forged cars (\S\ref{sec:method:lop}).

% \subsection{Limitations of Previous Defenses}
\subsection{Previous Defenses}
\label{sec:Limitation}

\noindent$\bullet$\textbf{ Rationale behind Defenses by Elimination}. To eliminate the forged vehicles crafted by appearing attacks, a defense would unavoidably remove a small ratio of detected real objects from the prediction of 3D object detectors. However, we argue this would hardly cause as substantial damages to the ADS as the \lyf{mistake} of detecting forged vehicles. It is mainly because: (i) As described in the attacker's goal, obstacles which appear near the ADS take the most decisive effect on the vehicle's future planning. Therefore, incorrect elimination of a real obstacle far from this vehicle may have limited influence on the decision-making of the ADS \cite{jiachen2020towards}. (ii) In ADS, the \textit{multi-object tracking} (MOT) module which follows the perception module will take the predictions from the LiDAR-based object detectors as input, maintain and predict the trajectories of objects nearby \cite{xinshuo2020AB3DMOT,hsu2021P3DMOT,wenhan2021motReview}. By design, MOT usually creates an object trajectory for a newly predicted object which is constantly detected for $6$ frames, while removes an overdue object trajectory which is continuously unmatched with any predicted objects for $60$ frames in a common visual perception system \cite{ji2018onlineMOT} of $30$ FPS. This mechanism guarantees that it is much easier for an ADS to create a fake object in its perception \lyf{due to} a successful appearing attack than forgetting a real object, due to the occasional misprediction of the LiDAR-based detector itself or the incorrect elimination of some real objects by such a defense.

Therefore, it is reasonable to tolerate a small ratio of false alarms from defenses by elimination and recognize the importance of defending against appearing attacks by slightly trading the recall of LiDAR-based object detectors. 
% For more background on MOT, please refer to Appendix \ref{appendix:mot}. 
However, the existing defense methods which are possibly against appearing attacks remain limitations in their design, so it is hard for them to maintain good performance in different scenes. To make it clear, we further analyze these defense methods and discuss their limitations accordingly.

\noindent$\bullet$\textbf{ Limitations of Universal Defenses.}
SRS (Simple Random Sampling) and SOR (Statistical Outlier Removal) are two universal defense methods for PC models. They are both unaware of attacks and against adversarial attacks by removing suspect points in input PC.

\noindent\textbf{(1) SRS.} SRS is in essence a random method regardless of any auxiliary information \cite{hang2019dupnet}. Formally speaking, given a raw input PC $X$ with $n$ points, SRS will randomly sample $M(M<n)$ points from $X$ by $
P(X)=\{\mathbb I_x|x\in X,\ \mathbb I_x\sim Bernoulli(0.5)\}, % \label{equal:bernoulli}
$
where $\mathbb I_x$ indicates the existence of each point $x$ in $X$.

\noindent\textbf{(2) SOR.} For a raw input PC $X$, SOR computes the average of the $k$-nearest neighbors' (kNN) distances for each point in $X$, and counts the mean $\mu$ and the standard deviation $\sigma$ of these distances. Then, it recognize those points which fall outside the range of $[\mu-\alpha\cdot\sigma,\mu+\alpha\cdot\sigma]$ as noises and removes them from $X$, where $\alpha=1.1$ is its hyper-parameter \cite{hang2019dupnet}.

\noindent$\bullet$\textbf{ Limitations of Specific Defenses.} 
CARLO (oCclusion-Aware hieRarchy anomaLy detectiOn), SVF (Sequential View Fusion) and Shadow-Catcher are three specific heuristic defense methods for 3D object detectors.
They both specify the attack as a black-box appearing attack proposed in Sun's work \cite{jiachen2020towards}, and perform defense by removing suspect points in input PC or deleting suspect objects in the final prediction.

\noindent\textbf{(1) CARLO.} CARLO is a heuristic defense algorithm proposed by Sun et al. to detect the cars forged by their black-box appearing attack \cite{jiachen2020towards}.
For each object predicted by the 3D object detectors, CARLO computes an anomalous ratio $r$ in one of the following two ways:
(1) \textbf{FSD (Free Space Detection)}, which defines $r=\sum_{c\in S^c}FC(c)/|S^c|$, where $S^c$ is a set including all the cells in this object's bounding box, and $FC(c)$ is a $0/1$ function indicating whether there are input points in the cell $c$; and (2) \textbf{LPD (Laser Penetration Detection)}, which defines $r=|{S\downarrow}^p|/|{S\downarrow}^p\bigcup S^p\bigcup {S\uparrow}^p|$, where the superscript $p$ indicates the corresponding set is composed of points. Specifically, ${S\downarrow}^p$ contains the input points in the space behind this object, $S^p$ is contains the input points inside this object's bounding box, and ${S\uparrow}^p$ contains the input points in the space between this object and the LiDAR.
% \begin{itemize}
%     \item \textbf{FSD (Free Space Detection)} In this way, the ratio $r=\frac{\sum_{c\in S^c}FC(c)}{|S^c|}$, where $S^c$ is a set including all the cells in this object's bounding box, and $FC(c)$ is a $0/1$ function indicating whether there are input points in the cell $c$.
%     \item \textbf{LPD (Laser Penetration Detection)} In this way, the ratio $r=\frac{|{S\downarrow}^p|}{|{S\downarrow}^p\bigcup S^p\bigcup {S\uparrow}^p|}$, where the superscript $p$ indicates the corresponding set is composed of points. Specifically, ${S\downarrow}^p$ contains the input points in the space behind this object, $S^p$ is contains the input points inside this object's bounding box, and ${S\uparrow}^p$ contains the input points in the space between this object and the LiDAR.
% \end{itemize}

Then, CARLO compares all these $r$ with a fixed threshold $R$. For those objects with $r>R$, CARLO recognizes them as fake objects and erases them from the prediction.

\noindent\textbf{(2) SVF.} Similarly, SVF is another defense algorithm suggested by Sun et al., but its key is more similar to SOR: removing outliers from the raw input PC.
As an extra end-to-end network, SVF turns the raw input PC into front-view (FV) representation and uses LU-Net\cite{pierre2019lunet}, a PC segmenter, to calculate a segmentation score for each point.
SVF then concatenates these scores with their related points' input features to re-generate the input PC, and passes this augmented PC to the 3D object detector as input.

\noindent\textbf{(3) Shadow-Catcher.} As our concurrent work, Shadow-Catcher \cite{zhongyuan2021shadowcatcher} also exploits the physical law to improve the robustness of the 3D object detectors in self-driving system. However, Shadow-Catcher is mainly based on hand-crafted rules to determine the forged obstacles, while our work proposes the first learning-based defense scheme to model the complicated physical relation between the depth and density of real objects for defensive purposes. Specifically, Shadow-Catcher computes an anomaly score for each detected object based on the distances of the points inside its bounding box to four key lines related to its bounding box, then compare this score with a presetting threshold to determine whether the perceived obstacle is forged. 
   
As a final remark, most of the previous defenses are initially designed for mitigating specific appearing attacks.
% , regardless of the existence of other potential attacks. 
In this sense, the performance of previous defenses against each popular attack remains unjustified in a systematic way, which we accomplish in our evaluation. 
\section{Defense with Local Objectness Predictor}\label{sec:OurDefense}
\begin{figure*}[htpb]
    \centering
    \includegraphics[width=0.95\textwidth]{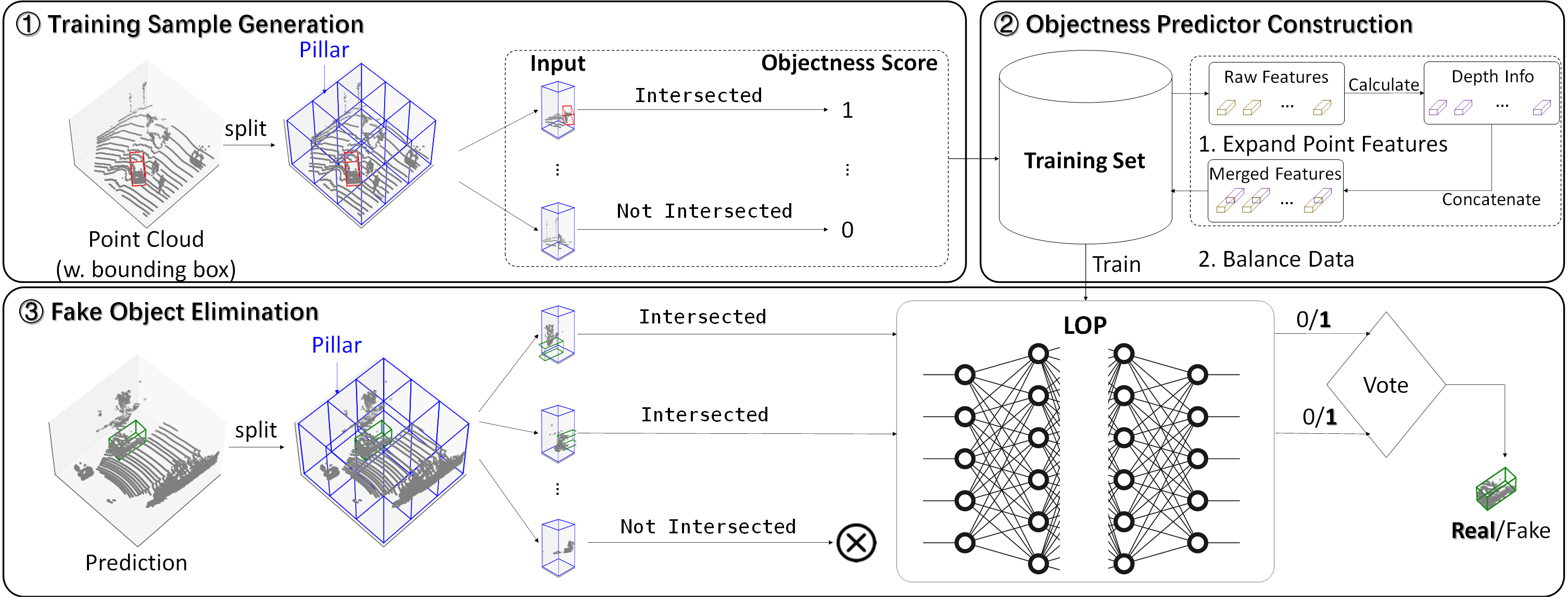}
    % \caption{The pipeline of our local objectness predictor. The red boxes represent the ground truths of original training sample, the green box represent the predicted objects of 3D object detectors, and the blue box represent the equal-size pillars separated from input space.}
    \caption{The pipeline of our proposed defense. The input space is split into a number of equal-sized pillars (in the form of blue boxes). The red box in \raisebox{.5pt}{\textcircled{\raisebox{-1pt} {1}}} represents the bounding box of a ground-truth \louis{object during training}, while the green box in \louis{\raisebox{.5pt}{\textcircled{\raisebox{-1pt} {3}}}} represents
    \louis{that of a predicted object from the 3D object detector during testing.}
    % the predicted bounding box of a test object from the 3D object detector.
    }
    \label{fig:local_objectness_detector}
\end{figure*}

% In this section, we propose a local objectness predictor \louis{(LOP)} to detect the fake objects forged by existing appearing attacks.

% \subsection{Defensive Insights.}
% Our defensive insights are derived from a closer look at the attack process of an appearing attack. 

% \noindent\textbf{Attack Process.} Before the attack starts, the attacker deploys the physical equipment shown in Fig.\ref{fig:appearing_attack_equipment}. Once the photodiode receives the lasers emitted by the victim autonomous vehicle's LiDAR, it activates the delay component, which then controls the laser transmitter to emit lasers to the LiDAR for spoofing. In other words, the delay component determines the travel of time of emitted lasers, the laser transmitter determines the light intensity of emitted lasers, and the lens determines the direction of emitted lasers, the attacker can inject a certain amount of points within a certain area as his/her wish. Later, the LiDAR-based 3D object detectors of the victim takes the infected PC and predicts an inexistent vehicle as the attacker wishes. 

% The process is full physical and is therefore restricted by physical laws. 
% % Finally, the victim changes its future travel plan to avoid the inexistent vehicle, \louis{which might lead to} the traffic jam or even crash accidents.

% As is shown 

\noindent\textbf{Methodology Overview.} As shown in Fig.\ref{fig:local_objectness_detector}, the pipeline of our proposed defense can be divided into three stages: training sample generation, objectness predictor construction and fake object elimination.
In the training sample generation stage, we construct a learning task for our local objectness predictor (LOP), which consists of pairs of \louis{points inside a small local pillar and its corresponding} objectness label, annotated in a fully self-supervised way without additional annotation except for a standard training dataset for LiDAR-based object detectors.
Then, in the objectness predictor construction stage, we train the LOP to learn to predict the objectness score \louis{for each pillar}, i.e., the confidence of whether a \louis{local part} belongs to a real object.
% the possibility that indicates the existence of a small part of real object, for input PC. 
Finally, in the fake object elimination stage, we use our trained LOP to predict an objectness score for \louis{each small pillar} intersected with the bounding \louis{boxes of the predicted objects,}
and determine whether these objects are real by majority voting.
Below, we elaborate on the insights and the technical designs in each stage of our defense.

\subsection{Training Sample Generation}
\label{sec:method:preparation}

\subsubsection{Insight: Global Objectness $\neq$ Local Objectness} 
By inspecting the design of recent appearing attacks, we observe that most attacks focus on increasing the confidence scores of the forged obstacles, which represents the possibility of the detected object to be real.
Equivalently, according to our definition of objectness, the confidence score can be explained as a \textit{global objectness score} related with the predicted obstacle to some extent.
As most LiDAR-based object detectors by design keep those objects with higher confidence, or global objectness scores, in their final predictions, increasing confidence scores is the most direct way for the attacker to successfully forge a non-existent obstacle. 
However, to increase the global objectness score of a forged obstacle does not necessarily lead to a higher objectness score for each local part.
With the following experiments, we observe that most of the recent appearing attacks have ignored 
\louis{the local difference, i.e. the spatial distance of two corresponding subsets, between a real and a forged obstacle,}
% the differences \louis{of local parts between the real and} their forged objects, 
which leaves an exploitable \louis{trace} for the defender.

\noindent$\bullet$\textbf{ A Pilot Study.} As the description in Section \ref{sec:ThreatModel}, the mainstream appearing attacks all focus on forging cars, so we mainly validate the above observation on cars.
We first randomly sample one real car
from the training set of KITTI \cite{andreas2012kitti} and $1,000$ forged cars crafted by three mainstream appearing attacks \cite{yulong2019advSensor, jiachen2020towards, kaichen2021robust} (later described in Section \ref{sec:Experiment}).
Next, we translate the interior points of each ground-truth car and each forged car into its local coordinate system, rotated by the lead angle to the identical orientation.
Then, for the point set $S$ of the real car and $S'$ of each forged car, we measure the distance between them by using the chamfer distance\cite{chong2019generating} and the average square L2 distance of kNN as metrics. 
% The details of these two popular metrics are shown in Appendix \ref{appendix:metrics}.
% We then calculate the differences on PC between the real vehicle and each forged vehicle with the following two different measurements:

\begin{figure}[t]
    \centering
    \includegraphics[width=0.45\textwidth]{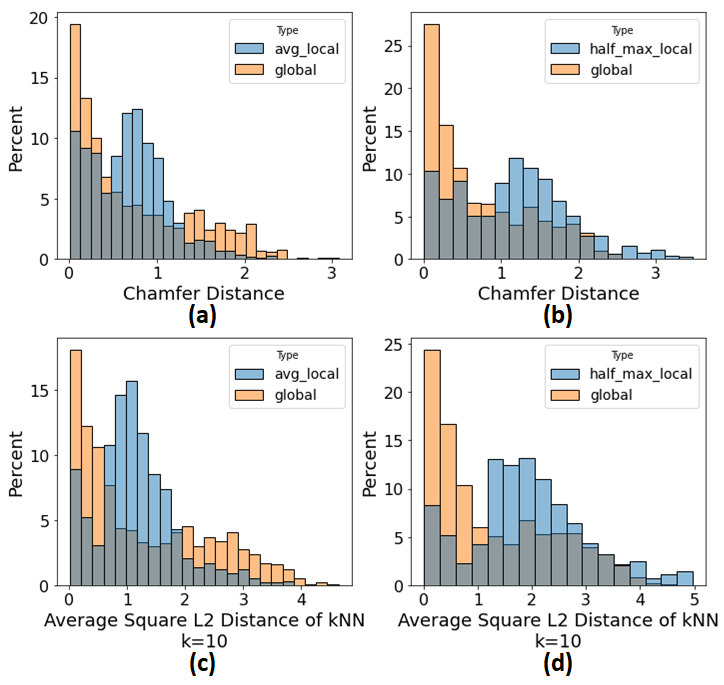}
    \caption{The local and global differences \louis{of PCs between} real and forged cars (\scriptsize{The grey bars inside denote the overlapping region.}).}
    \label{fig:local_global}
\end{figure}

Specifically, we collect points belonging to the real car as $S_R$. For each forged car, we first collect points belonging to it as $S_F$.
Then, we split the point space into equal-sized pillars $p_j$ (as in Fig.\ref{fig:local_objectness_detector}), and generate \louis{a point subset} $S_{F,j}= S_F \cap p_j$ for each pillar.
Finally, we calculate the global difference and local difference as follows:
\begin{equation}
    D_{\text{global}} = D(S_R,S_F) \label{equal:global}
    % & D_{max\_local} = max\{D(S_R,S_{F, j})\} & \label{equal:max_local}
\end{equation}
\begin{equation}
    D_{\text{avg\_local}} = \frac{1}{|\{S_{F,j}\}|}\sum D(S_R,S_{F,j}) \label{equal:avg_local}
\end{equation}
\begin{equation}
    D_{\text{half\_max\_local}} = \frac{1}{N_{\text{half}}}\text{Top}_{N_{\text{half}}}(\{D(S_R,S_{F,j})\}) \label{equal:half_max_local}
\end{equation}
where $S_R,S_F$ denote the two specific point sets defined above, $S_{F,j}$ denotes the point sets gathered from the separated pillars of $S_F$, $D\in\{D_C,D_k\}$ denotes the metric that we use to measure distance between two point sets,
$N_{\text{half}} = \lceil |\{S_{F,j}\}|/2 \rceil$ is half of the number of point subsets $S_{F,j}$,
and $\text{Top}_{k}(V)$ denotes the sum of the largest $k$ values in $V$.
% and $Top_{cnt}(V)$ is a function that calculate the sum of first $cnt$ large value in $V$.

As shown in Fig.\ref{fig:local_global}, the local differences of the forged cars are usually larger than the global differences in both
% two different metrics above
chamfer distance and average \louis{square} L2 distance of kNN 
(with all p-value less than $1.0\times 10^{-11}$ in Kolmogorov-Smirnov tests).
We further compare
the local difference and global difference
% $D_{\text{local}}$ and $D_{\text{global}}$
for each forged car, and find that if we choose $D_{\text{avg\_local}}$ as the local difference, there are $55.7\%$ forged cars have larger local difference on the chamfer distance metric, and $54.5\%$ forged cars have larger local difference on the average \louis{square} L2 distance of kNN metric.
If we choose $D_{\text{half\_max\_local}}$ as the local difference, $87.4\%$ forged cars have larger local difference on the chamfer distance metric, and $87.5\%$ forged cars have larger local difference on the average \louis{square} L2 distance of kNN metric.
\louis{Similar results are observed when we repeat the experiment above on several other real cars randomly sampled from the training set of KITTI.}

In summary, the experimental results imply that \textit{the local features do provide the defender with a \louis{trace} to distinguish between the real and forged cars.}
In fact, our insight also conforms to a recent work on enhancing the precision of LiDAR-based object detectors \cite{qi2020object}, where they suggest that with an appropriate strategy of spatial division, one small part of real objects can also contain rich enough spatial and semantic information to predict the category, bounding box and confidence score of its related object. 

% [Kolmogorov-Smirnov Test Result]
% avg_local:
% KstestResult(statistic=0.164, pvalue=1.8715496208415e-12)
% KstestResult(statistic=0.384, pvalue=2.0919998989157213e-66)
% half_max_local:
% KstestResult(statistic=0.307, pvalue=2.6373240994946432e-42)
% KstestResult(statistic=0.544, pvalue=2.0738160340030527e-136)

\subsubsection{Technical Designs} 
% \noindent$\bullet$ \textbf{ Technical Designs.} 
To facilitate the modeling of local object features, in the first stage we prepare a dataset ${D}_\text{obj}$ consisting of pairs of \louis{points in each pillar} from ground-truth objects \louis{and} an automatically annotated objectness label based on a standard training dataset for LiDAR-based object detectors (e.g., KITTI \cite{andreas2012kitti}).
Formally, we denote the training dataset as $D = \{(X_t, \{\mathbf{b}_k\}_{k=1}^{N_t})\}_{t=1}^{N}$, where $N_t$ denotes the number of ground-truth objects in the PC $X_t$, and $\mathbf{b}_k$ denotes the bounding box of the $k$-th ground-truth object in $X_t$.
First, we split the full $L\times W\times H$ 3D region which covers the input point clouds into a number of pillars $\{p_j\}$ with an equal size $l \times {w} \times {H}$, where $l=1m,w=1m$ in our implementation.
Then for each pillar $p_j$, we generate an input-output pair, which can be represented as $(pc_j,obj_j)$, as follows:

\noindent\textbf{Generating Input $pc_j$.} We directly collect the inside points of each pillar from the input PC $X_t$ to form the input feature $pc_j$, i.e., $pc_j = X_t \cap {p_j}$, \louis{composed of a batch of points' features $x_i$ inside $p_j$}.
To normalize the generated input, we constrain the size of $pc_j$ as $M_{pc}$, where $M_{pc}$ is a fixed hyper-parameter. For those $pc_j$ with a larger size, we randomly sample $M_{pc}$ interior points as its input. Otherwise, $pc_j$ is padded with $\vec{0}$ until the size constraint is satisfied.

\noindent\textbf{Generating Label $obj_j$.} We first calculate the 2D Intersection over Union (IoU), the ratio of the area of intersection region over that of union region, between $p_j$ and each ground-truth bounding box $\mathbf{b}_k$ on the x-y plane.
\louis{For each pillar $p_j$,} we keep the maximal IoU value over all ground-truth bounding boxes.
Finally, we compare the maximal IoU value with a fixed threshold $T_{\text{IoU}}$.
If this value is greater than $T_{\text{IoU}}$, we annotate $obj_j = 1$ to indicate that the pillar $p_j$ contains a local part of a real object,
or \louis{$obj_j=0$ otherwise.}
Iterating over all the PC inputs with the pillars, we finish the collection of the training set $D_\text{obj} = \{(pc_j, obj_j)\}$.
As an analogy to the training task of masked word prediction for pretrained language models \cite{jacob2019bert}, this process works in a fully self-supervised manner without any additional information.

\subsection{Objectness Predictor Construction}
\label{sec:method:lop}

\subsubsection{Insight: The Inimitable Depth-Density Law}
Meanwhile, we find that, because recent appearing attacks are designed to cause threats in the real world, they are inevitably limited by certain physical constraints imposed by both the attacker's goal and the attack apparatus.
As is introduced in Section \ref{sec:Limitation}, there exist physical upper bounds on the number of added points and the permissible distance between a fake object and LiDAR for recent appearing attacks.
Behind these two limitations, we find that the capability of recent appearing attacks is inherently restricted by the \textit{depth-density} law \cite{depth_density}:
with existing technology and methods, it is hard to imitate \louis{the real-world objects' relation} between the \textit{depth}, i.e. the distance between this object and the LiDAR, and the \textit{point density}, i.e. the ratio of the number of input points inside this object's bounding box over the volume of its bounding box.

\begin{figure}[ht]
    \centering
    \includegraphics[width=0.45\textwidth]{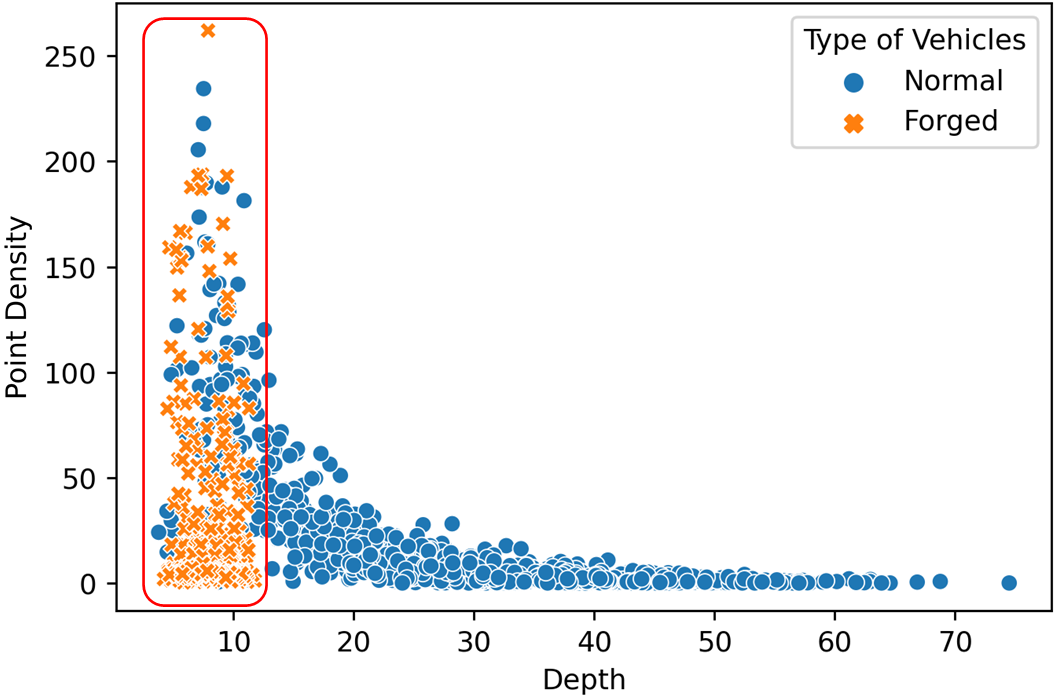}
    \caption{The distribution chart of the depth-density relation.
    \lyf{The blue points represent normal cars, the orange crosses represent forged cars, and the red rectangle shows the confounding region of the two.}
    }
    \label{fig:depth_density}
\end{figure}

\noindent$\bullet$\textbf{ A Pilot Study.} 
Similar to the reason introduced in Section \ref{sec:method:preparation}, we mainly validate the above observation on cars here.
We first randomly sample $1,000$ real cars from the training set of KITTI and $1,000$ forged cars crafted by the mainstream appearing attacks described in Section \ref{sec:Experiment}.
Then, we calculate the depth and point density for these objects based on their bounding boxes and the related points.
As shown in Fig.\ref{fig:depth_density}, the point density of real cars is approximately inversely proportional to their depth.
In contrast, the point density of the forged cars seems to be independent of the depth: they can have small depth and small point density simultaneously, while this seldom happens for real cars. 

Though differences exist between real and forged cars in terms of the depth-density relation, it is still hard to directly distinguish them by heuristic algorithms.
Due to the complexity of real-world environments, there exists the confounding region in the depth-density relation \louis{distribution} (highlighted in Fig.\ref{fig:depth_density}), which is mainly caused by some real cars occluded by others, with smaller depth and point density at the same time.
Besides, the complexity is further increased by errors such as the noise in LiDAR perception and the deficiency of attack equipment.
In other words, it can be improper to explicitly filter out any detected object based on the hand-crafted rules.
As a data-driven approach, we alternatively \louis{encourage} the LOP to actively learn to model the complicated depth-density relation of real objects, by further incorporating the depth information explicitly into the input feature of each pillar we derive in the first stage.

\subsubsection{Technical Designs}
At this stage, we augment the input features in our prepared training dataset $D_\text{obj}$ with the depth information.
Specifically, for each generated training sample $(pc_j, obj_j)$ in $D_\text{obj}$,
\louis{we expand the feature of each point in $pc_j$ from an original $4$-dim vector $x_i = (x,y,z,int)$ into a $7$-dim one $x_i' = (dx,dy,x,y,z,int,dep)$, where}
\louis{% $(x,y,z)$ is the point's 3D coordinates,
$(dx,dy)$ is the point's 2D relative coordinates to the center of its corresponding pillar in the x-y plane,
% $int$ is the point's intensity
and $dep=\sqrt{x^2+y^2+z^2}$ is the point's depth.} In our preliminary, we also experimented with an alternative design with no depth information explicit in the input feature. The practice would result in a LOP which is much less effective in distinguishing the forged objects from the real ones than using our current solution.

% \noindent$\bullet$ \textbf{ Technical Designs.} At this stage, we augment the input feature in our prepared training dataset $D_\text{obj}$ with the depth information. Specifically, for each generated training sample $(x_j, y_j)$ in $D_\text{obj}$, we expand the feature vector of each point in $x_j$ from $4$-dim feature vector $(x,y,z,int)$ into a $7$-dim feature vector $(dx,dy,x,y,z,int,dep)$, where $(x,y,z)$ represents the point's 3D coordinates, $(dx,dy)$ represents the point's 2D relative coordinates to the center of its corresponding pillar in x-y plane, $int$ represents the point's intensity and $dep=\sqrt{x^2+y^2+z^2}$ represents the point's depth.

To adaptively learn the depth-density relation for distinguishing real and forged cars or other obstacles, we implement the LOP $O$ with the architecture of an off-the-shelf backbone PC classifier (e.g., PointNet \cite{charles2017PN} or DGCNN \cite{yue2019DGCNN}),
considering their validated performance on many downstream 3D tasks.
Note that the \textit{negative samples} in $D_\text{obj}$, i.e. the generated samples with $obj_j=0$, are much more than the \textit{positive samples}, i.e. the generated samples with $obj_j=1$.
Thus, we delete a part of negative samples in random to keep data balance and ensure that the ratio of positive samples and negative samples does not exceed $1:1.5$.
To further alleviate the data imbalance problem, we also adopt the idea of focal loss \cite{tsungyi2017focalloss} in the learning objective of LOP:
\begin{align}
FL(p,y)=-\alpha_{fl}(1-p_y)^{\gamma_{fl}} log(p_y) \label{equal:focalLoss}
\end{align}
where the positive constants $\alpha_{fl},\gamma_{fl}$ ($\gamma_{fl}>1$) are the hyper-parameters of the focal loss, which are set by following the best practices in \cite{tsungyi2017focalloss}. Besides, $p_y$ is the probability of the $y$-th class returned by the predictor.

%%%%%%% BEGIN ASR-Precision Map
\begin{figure}[t]
    \centering
    \includegraphics[width=0.45\textwidth]{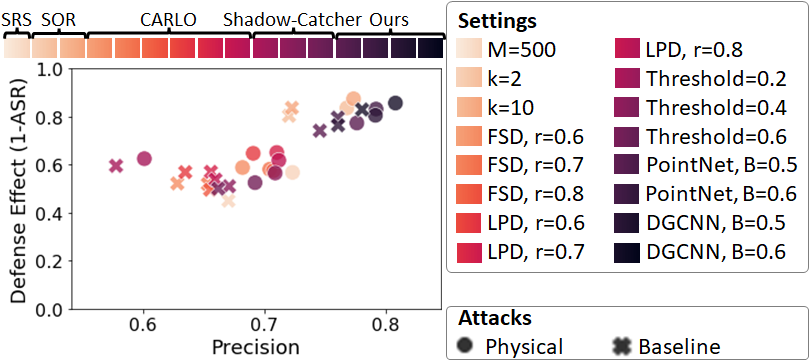}
    \caption{The relation graph of defense effect (1-ASR) and precision on cars of PointPillars under attacks. 
    % \lyf{The defenses are represented in different colors, while the attacks are represented in different shapes.}
    "PointNet" and "DGCNN" refers to LOP's structure, \louis{with a boundary value $B$} used to distinguish real and fake objects as the description in Section \ref{sec:method:voting}."LPD" and "FSD" are two strategies for CARLO to calculate the anomalous ratio, and $M,\ k,\ R$, Threshold are the hyper-parameters of other defenses, which are all described in Section \ref{sec:Limitation}.
    }
    \label{fig:comp_atk_pre_asr}
\end{figure}
\begin{figure}[ht]
    \centering
    \includegraphics[width=0.45\textwidth]{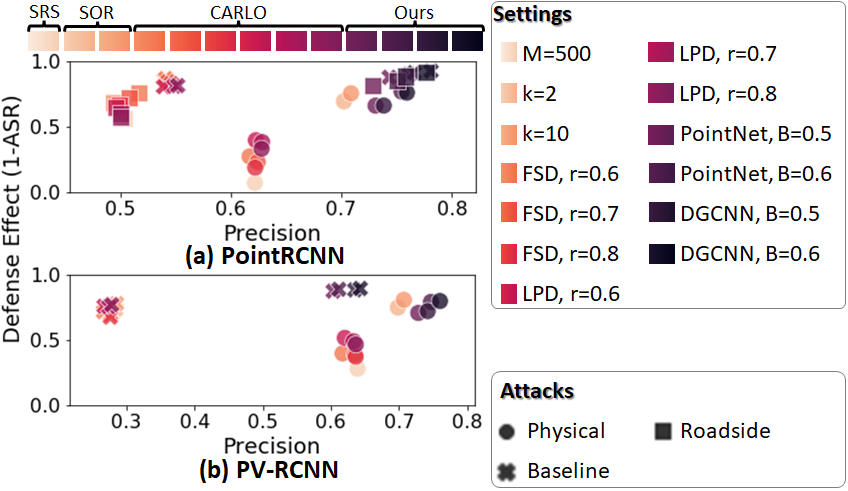}
    \caption{The relation graph of defense effect and precision on cars of (a) PointRCNN and (b) PV-RCNN under attacks.
    }
    \label{fig:comp_atk_pre_asr_0}
\end{figure}
%%%%%%%% END ASR-Precision Map

\subsection{Fake Object Elimination}
\label{sec:method:voting}
Finally, we leverage the LOP to calculate the objectness score for each pillar intersected with predicted objects, and determine whether these objects are real by a majority voting among the pillars. Specifically, we first divide the detection space into
% We first divide the detection space into
\louis{equal-sized pillars, translate the input PC into a series of point subsets inside these pillars and then augment their features, similarly to the former two stage.}

% turn them into input PCs similarly to the sample generating stage and the training stage.
Then we use the LOP to calculate a $0/1$ objectness score for each pillar.
%%%% Technically speaking, these objectness scores are shared among all the predicted objects.
For each object in the prediction of the 3D object detector,
we search for those pillars \louis{whose} 2D IoU on the x-y plane between itself and the predicted object's bounding box is greater than a specified threshold 
$\beta$,
and calculate the sum of their objectness scores \louis{as well as the ratio of this sum over the total number of related pillars.}
Finally, we recognize those objects with the ratio
% of this sum over the number of counted pillars
less than or equal to a boundary value $B$ as fake objects, and eliminate them from the prediction. 
% For more details, please refer to Algorithm \ref{alg:detecting} in Appendix \ref{appendix:FOE}. 
% The pseudo-code of this stage is described in Appendix \ref{appendix:stage3_code}.
\section{Evaluation and Analysis} \label{sec:Experiment}
% In this section, we deploy our LOP on three state-of-the-art 3D object detectors to evaluate their performances in normal circumstances \louis{as well as} under appearing attacks. 
% Besides, we further compare the defense effect of our LOP with other existing defenses against appearing attacks. Below, we describe the setup of experiments and report the results.

\subsection{Overview of Evaluation}\label{sec:Experiment:setup}
\noindent\textbf{(1) Victim Models.} We choose three mainstream LiDAR-based object detectors: \textbf{PointPillars} \cite{alex2019pointpillars}, \textbf{PointRCNN} \cite{shaoshuai2019pointrcnn} and \textbf{PV-RCNN} \cite{shaoshuai2020pvrcnn} as the victim models. Specifically, we adopt the implementation of these three object detectors available on an open-source project OpenPCDet \cite{OpenPCDet}, each of which is normally trained on KITTI dataset \cite{andreas2012kitti} to achieve the near state-of-the-art performance. 

% OpenPCDet is an open-source testing platform for 3D object detection.  % As the 3D object detectors may predict some objects with a low confidence score in prediction, we follow the convention in \cite{joseph2016yolo,shaoshuai2019pointrcnn,shaoshuai2020pvrcnn} and only consider the objects with confidence scores larger than a threshold $C$ in their prediction to keep their precision at a normal level. 

\noindent\textbf{(2) Attack Methods.} We implement three popular appearing attacks which can be roughly categorized into \textit{white-box} and \textit{black-box} attacks. In the former case, the attacker has full access to the victim 3D detector, including the architecture and the parameters, while the latter only accesses the detector as a black-box prediction API. Specifically, the attacks are
    
    \noindent \textbullet \textit{\ A Variant of Adv-LiDAR} \cite{yulong2019advSensor}  (\textit{abbrev.} \textbf{Baseline}, \textit{white-box}): Because Adv-LiDAR is specially designed for attacking Baidu's Apollo \cite{Apollo}, it could hardly be directly transferred to attack other 3D object detectors \cite{jiachen2020towards}. Therefore, following its main idea, we implement a variant of ADV-LiDAR by randomly injecting a certain number of points into a specified zone, and using FGSM \cite{ian2015explaining} to increase the confidence scores of those forged cars related to these points.

    \noindent \textbullet \textit{\ Yang's Work} \cite{kaichen2021robust}  (\textit{abbrev.} \textbf{Roadside}, \textit{white-box}): This attack forges cars by 3D printing a small and specifically-designed object, increasing their confidence scores and category scores of label car, and enlarging their bounding boxes with gradient descent. 
    \xqf{Since this attack will generate the adversarial points and then turn them into a physical object, we will only deploy the first part for our experiments.}
    In their original work, this attack mainly aims at breaking PointRCNN with a white-box access. We thus also follow the settings in our experiments.
    
   \noindent \textbullet \textit{\ Sun's Work} \cite{jiachen2020towards}  (\textit{abbrev.} \textbf{Physical}, \textit{black-box}): This attack forges cars by duplicating real cars, which contains a limited number of points due to either inter-occlusion or intra-occlusion. The PC of the fake car is then transformed to a front-near position of the victim ADS.

\noindent\textbf{(3) Baseline Defenses.} We implement \textbf{SRS}, \textbf{SOR}, \textbf{CARLO} and \textbf{Shadow-Catcher} which we have introduced in Section \ref{sec:Limitation} as the baseline defenses.
We do not consider SVF because it relies on retraining the whole 3D object detector itself, and causes much more time and computation cost compared with other baseline defenses as well as ours (Section \ref{sec:Experiment:results:overhead}).
% Also, as CARLO and SVF are specially designed for the same attack in \cite{jiachen2020towards}, we argue that the results of CARLO can partly represent that of SVF.

\noindent\textbf{(4) Metrics.}
We choose three different metrics to evaluate the performance of our proposed defense and other defenses:

    \noindent{\textbullet\ \textbf{Precision}} measures the proportion of the real objects in the prediction results. In the context of defense, the decrease in precision reflects whether these defenses would harm the original performance of the victim model.
    Following  \cite{alex2019pointpillars,shaoshuai2019pointrcnn,shaoshuai2020pvrcnn}, we first choose a certain threshold $C_{\text{conf}}$ for each 3D object detector, and remove those predicted objects with confidence scores less than $C_{\text{conf}}$. 
    Then we calculate the 2D IoU on the x-y plane between the bounding boxes of each remaining predicted object and the ground-truth objects. A predicted object is true positive, if its maximal IoU value surpasses another certain threshold $C_{\text{IoU}}$ and the predicted category is identical with the ground-truth; otherwise, the predicted object is a false positive prediction.
    For PointPillars and PointRCNN, we set $C_{\text{conf}}=0.5,\ C_{\text{IoU}}=0.5$; for PV-RCNN, we set $C_{\text{conf}}=0.7,\ C_{\text{IoU}}=0.5$.

    \noindent{\textbullet\ \textbf{Average Precision (AP)}} is a comprehensive metric over the precision and the recall of the detection results. Specifically, AP is the average value of precision when the recall is larger than different specific values, which can be represented as
\begin{align}
AP=\frac{1}{11}\sum_{r\in\{0,0.1,\hdots,1.0\}}{max}_{r'\geq r}\text{Precision}@(\text{Recall}= r')
\end{align}

    \noindent{\textbullet\ \textbf{Attack Success Rate (ASR)}} measures the ratio of the number of forged cars detected by the victim 3D object detector over the total number of attack attempts, which directly reflects the performance of these defenses. A more effective defense should result in a lower ASR. 

\noindent\textbf{(5) Implementation of LOP.} We choose two off-the-shelf point-wise PC classification architectures, PointNet \cite{charles2017PN} and DGCNN \cite{yue2019DGCNN}, to instantiate the LOP. For those hyper-parameters of LOP described in Section \ref{sec:OurDefense}, we set $M_{pc}=1024,\ T_{\text{IoU}}=1\times{10}^{-6},\ \alpha_{fl}=1,\ \gamma_{fl}=2,\ \beta=1\times{10}^{-3}$.

\begin{table*}[htbp]
  \centering
  \caption{The AP on cars of PointPillars with and without LOP or other defense methods under attacks.}
\scalebox{0.75}{
    \begin{tabular}{lccccccccc}
    \toprule
          & \multirow{2}[0]{*}{None} & \multicolumn{2}{c}{SOR} & \multicolumn{6}{c}{CARLO} \\
    \cmidrule(lr){3-4} \cmidrule(lr){5-10} 
          &       & k=2   & k=10  & FSD, r=0.6 & FSD, r=0.7 & FSD, r=0.8 & LPD, r=0.6 & LPD, r=0.7 & LPD, r=0.8 \\
    \midrule
    Physical & 70.06\% & 70.43\% & 70.01\% & 65.94\% & 69.60\% & 69.64\% & 67.79\% & 69.57\% & 69.82\% \\
    Baseline & 68.57\% & 68.84\% & 68.06\% & 63.49\% & 67.96\% & 67.96\% & 64.37\% & 67.99\% & 68.49\% \\
    \toprule \bottomrule
          & \multirow{2}[0]{*}{None} & SRS   & \multicolumn{3}{c}{Shadow-Catcher} & \multicolumn{4}{c}{Ours} \\
    \cmidrule(lr){3-3} \cmidrule(lr){4-6} \cmidrule(lr){7-10}
          &       & M=500 & Threshold=0.2 & Threshold=0.4 & Threshold=0.6 & PointNet, B=0.5 & PointNet, B=0.6 & DGCNN, B=0.5 & DGCNN, B=0.6 \\
    \midrule
    Physical & 70.06\% & 70.12\% & 47.72\% & 75.46\% & 77.05\% & 70.92\% & 70.97\% & 71.11\% & 70.39\% \\
    Baseline & 68.57\% & 68.55\% & 57.81\% & 75.03\% & 75.95\% & 69.50\% & 69.63\% & 69.72\% & 68.61\% \\
    \bottomrule
    \end{tabular}%
}
%   \vspace{-0.1in}
  \label{tab:comp_map_atk_0}%
\end{table*}%

\begin{table}[ht]
  \centering
  % \vspace{-0.2in}
  \caption{The AP on cars of PointRCNN and PV-RCNN with and without LOP or other defense methods under attacks.
%   "PointNet" and "DGCNN" refers to the model structure of LOP, \louis{with a boundary value $B$} used to distinguish real objects and fake objects as the description in Section \ref{sec:method:voting}."LPD" and "FSD" are two different strategies for CARLO to calculate the anomalous ratio, and $M,\ k,\ r$ are the hyper-parameters of other defenses, which are all described in Section \ref{sec:Limitation}.
  }
\scalebox{0.7}{
    \begin{tabular}{lccccc}
    \toprule
        & \multicolumn{3}{c}{PointRCNN} & \multicolumn{2}{c}{PV-RCNN} \\
\cmidrule(lr){2-4} \cmidrule(lr){5-6}          
        & Physical & Baseline  & Roadside & Physical & Baseline \\
    \midrule
    w/o. defense  & 67.92\% & 65.95\% & 61.59\% & 70.11\% & 66.39\% \\
    SRS (M=500) & 69.42\% & 65.44\% & 62.48\% & 70.14\% & 66.27\% \\
    \midrule
    SOR (k=2) & 72.56\% & 65.41\% & 60.84\% & 71.43\% & 64.80\% \\
    SOR (k=10) & 72.63\% & 65.26\% & 60.53\% & 71.28\% & 64.09\% \\
    \midrule
    CARLO(FSD, R=0.6) & 67.16\% & 63.53\% & 60.16\% & 67.99\% & 63.27\% \\
    CARLO(FSD, R=0.7) & 68.41\% & 65.29\% & 60.71\% & 70.10\% & 66.07\% \\
    CARLO(FSD, R=0.8) & 68.15\% & 65.22\% & 60.69\% & 70.15\% & 65.84\% \\
    CARLO(LPD, R=0.6) & 68.98\% & 65.27\% & 60.53\% & 69.23\% & 64.46\% \\
    CARLO(LPD, R=0.7) & 69.22\% & 65.97\% & 61.41\% & 70.26\% & 65.71\% \\
    CARLO(LPD, R=0.8) & 69.15\% & 66.04\% & 61.65\% & 70.35\% & 66.11\% \\
    \midrule
    Ours(PointNet, B=0.5) & 73.32\% & 71.87\% & 70.82\% & 71.47\% & 69.25\% \\
    Ours(PointNet, B=0.6) & \textbf{73.77\%} & 72.30\% & 71.41\% & 71.86\% & 69.19\% \\
    Ours(DGCNN, B=0.5) & 73.07\% & 72.29\% & 71.99\% & \textbf{71.87\%} & \textbf{69.88\%} \\
    Ours(DGCNN, B=0.6) & 73.74\% & \textbf{73.42\%} & \textbf{72.84\%} & 71.51\% & 68.30\% \\
    \bottomrule
    \end{tabular}%
}
%   \vspace{-0.1in}
  \label{tab:comp_map_atk}%
\end{table}%

\subsection{Comparison with Baselines}
\subsubsection{Attack Scenarios}
\label{sec:Experiment:results:comp_attack}
First, we evaluate the performance of our defense against \xqf{recent} appearing attacks. We implement three recent appearing attacks to generate adversarial examples against the three mainstream 3D object detectors based on the KITTI's validation set. We evaluate the ASR of these appearing attacks along with the AP and the precision of these detectors under attacks.
Besides the forged cars crafted by appearing attacks, there also remains some normal objects in the adversarial examples, which are considered in the AP and precision metrics.
Table \ref{tab:comp_map_atk_0} and Table \ref{tab:comp_map_atk} show the AP of the detectors on cars when equipped with different defenses,
and Fig.\ref{fig:comp_atk_pre_asr} and Fig.\ref{fig:comp_atk_pre_asr_0} plots the defense effectiveness (y-axis, in terms of $1-\text{ASR}$) and the precision on cars (x-axis) of different defenses.

\noindent\textbf{Results \& Analysis.} As we can see from Table \ref{tab:comp_map_atk_0}, Table \ref{tab:comp_map_atk}, Fig.\ref{fig:comp_atk_pre_asr} and Fig.\ref{fig:comp_atk_pre_asr_0}, compared with SRS and CARLO, our defense simultaneously achieves higher defense effectiveness and the victim models under guard have higher AP and precision on cars. 
For example, under \xqf{recent} appearing attacks, the PointRCNN equipped with the LOP keeps AP on cars over $70\%$ and precision on cars over $72\%$, while AP on cars is always less than $70\%$ and precision on cars is always less than $63\%$ when deploying SRS or CARLO on the PointRCNN.
Compared with SOR, although in some cases our defense has slightly lower defense effectiveness (the margin is less than $5\%$), it always results in higher AP and precision on cars under attacks. Compared with Shadow-Catcher, although in some cases our defense has slightly lower AP on cars, it always results in higher precision on cars and better defense effectiveness under attacks.

% In summary, our defense is effective against existing appearing attacks and incurs only slightly decrease on the normal performance of the victim detectors.

From a different perspective, we observe that other defenses only perform well when protecting certain models against specific attack techniques. For example,
SOR performs better when protecting PointPillars, while CARLO performs better when defending against \textit{Physical}. In contrast, the LOP performs well independent of the structure of the 3D object detector or the undergoing appearing attack, which implies that our proposed defense is more general than other defenses.

\subsubsection{Benign Scenarios}
\label{sec:Experiment:results:normal}
Then, we evaluate the performance of the victim models under guard on clean samples to measure the performance overhead brought by different defenses.
Table \ref{tab:comp_normal_pre_0} and Table \ref{tab:comp_normal_pre} present the AP and precision of them in the normal circumstances.

\begin{table*}[ht]
  \centering
  % \vspace{-0.2in}
  \caption{The AP and precision of PointPillars on cars with different defenses on clean samples.}
\scalebox{0.75}{
    \begin{tabular}{cccccccccc}
    \toprule
          & \multirow{2}[0]{*}{None} & \multicolumn{2}{c}{SOR} & \multicolumn{6}{c}{CARLO} \\
    \cmidrule(lr){3-4} \cmidrule(lr){5-10} 
          &       & k=2   & k=10  & FSD, r=0.6 & FSD, r=0.7 & FSD, r=0.8 & LPD, r=0.6 & LPD, r=0.7 & LPD, r=0.8 \\
    \midrule
    AP    & 72.34\% & 71.20\% & 70.58\% & 67.92\% & 72.03\% & 71.95\% & 69.80\% & 71.58\% & 72.02\% \\
    Precision & 78.99\% & 78.91\% & 78.86\% & 75.06\% & 77.81\% & 78.00\% & 75.77\% & 77.56\% & 78.31\% \\
    \toprule \bottomrule
          & \multirow{2}[0]{*}{None} & SRS   & \multicolumn{3}{c}{Shadow-Catcher} & \multicolumn{4}{c}{Ours} \\
    \cmidrule(lr){3-3} \cmidrule(lr){4-6} \cmidrule(lr){7-10}
          &       & M=500 & Threshold=0.2 & Threshold=0.4 & Threshold=0.6 & PointNet, B=0.5 & PointNet, B=0.6 & DGCNN, B=0.5 & DGCNN, B=0.6 \\
    \midrule
    AP    & 72.34\% & 72.33\% & 50.58\% & 77.41\% & 79.47\% & 72.86\% & 72.88\% & 73.63\% & 72.73\% \\
    Precision & 78.99\% & 79.14\% & 70.25\% & 77.31\% & 76.91\% & 81.77\% & 82.38\% & 83.04\% & 83.90\% \\
    \bottomrule
    \end{tabular}%
}
%   \vspace{-0.1in}
  \label{tab:comp_normal_pre_0}%
\end{table*}%

\begin{table}[ht]
  \centering
  % \vspace{-0.2in}
  \caption{The AP and precision of PointRCNN and PV-RCNN on cars with different defenses on clean samples.}
\scalebox{0.8}{
    \begin{tabular}{lcccc}
    \toprule
          & \multicolumn{2}{c}{PointRCNN} & \multicolumn{2}{c}{PV-RCNN} \\
    \cmidrule(lr){2-3} \cmidrule(lr){4-5}          
          & AP    & Precision & AP    & Precision \\
    \midrule
    w/o. defense  & 75.13\% & 75.04\% & 73.32\% & 73.12\% \\
    SRS (M=500) & 75.52\% & 74.75\% & 73.48\% & 73.23\% \\
    \midrule
    SOR (k=2) & 74.46\% & 74.49\% & 72.52\% & 72.92\% \\
    SOR (k=10) & 74.04\% & 73.88\% &72.22\% & 72.94\% \\
    \midrule
    CARLO(LPD, R=0.6) & 73.49\% & 73.10\% & 71.43\% & 70.08\% \\
    CARLO(LPD, R=0.7) & 74.63\% & 74.35\% & 73.21\% & 72.22\% \\
    CARLO(LPD, R=0.8) & 74.53\% & 74.32\% & 73.20\% & 72.41\% \\
    CARLO(FSD, R=0.6) & 74.17\% & 73.07\% & 72.00\% & 69.93\% \\
    CARLO(FSD, R=0.7) & 74.79\% & 74.38\% & 72.94\% & 71.66\% \\
    CARLO(FSD, R=0.8) & 74.89\% & 74.87\% & 73.15\% & 72.37\% \\
    \midrule
    Ours(PointNet, B=0.5) & 76.49\% & 79.29\% & 73.65\% & 77.85\%  \\
    Ours(PointNet, B=0.6) & 76.37\% & 80.03\% & 73.80\% & 78.53\% \\
    Ours(DGCNN, B=0.5) & 76.77\% & 80.75\% & \textbf{74.50\%} & 79.61\% \\
    Ours(DGCNN, B=0.6) & \textbf{76.84\%} & \textbf{81.52\%} & 73.86\% & \textbf{80.34\%} \\
    \bottomrule
    \end{tabular}%
}
%   \vspace{-0.1in}
  \label{tab:comp_normal_pre}%
\end{table}%

%  As is observed, the improvements is substantially larger than the improvement brought by other defenses.

\noindent\textbf{Results \& Analysis.} As Table \ref{tab:comp_normal_pre_0} and Table \ref{tab:comp_normal_pre} show, compared with existing defenses, the performance of 
% the three victim models 
these detectors
has less degradation in the normal cases when equipped with the LOP than with other defenses. For example, the AP and precision of 
% 3D object detectors 
detectors
equipped with other defenses both decrease in most cases, while for these 3D object detectors equipped with the LOP, the AP on cars even increases by $0.33\sim1.71\%$, and the precision on cars increases by $2.78\sim7.12\%$. Although Shadow-Catcher has slightly higher AP on cars than the LOP, 
% the detectors with Shadow-Catcher usually has lower precision than those detectors with LOP. Besides, 
considering the defensive advantages of LOP under different appearing attacks, our proposed defense may be more suitable for practical ADS, due to the performance-robustness balance when the detector is equipped with LOP.
% the choice of its threshold will significantly affect the performance of Shadow-Catcher and lead to a large trade-off between the performance and the robustness of the detector deployed with it, while \cite{zhongyuan2021shadowcatcher} only suggest to choose the value of it based on experience, which is harder to implement in the real world.}

% Although performance degradation is observed for the precision on pedestrians, the degradation is controlled below $2.2\%$. We infer the main reason of degradation is two-fold:
% On the one hand, the performance of the victim models on pedestrians is usually worse than that on cars and cyclists, which is a long-standing pain point for all existing object detectors \cite{alex2019pointpillars,shaoshuai2019pointrcnn,shaoshuai2020pvrcnn}.
% Therefore, it is also hard for the LOP to precisely detect the fake pedestrian based on the local pillars.
% On the other hand, the point density and the depth of pedestrians are usually smaller than cars and cyclists, which brings additional challenges in modeling the depth-density relation, a key to the performance of LOP. 

We further analyze why our proposed defense would even increase the performance of the victim models on cars in normal cases, while existing defenses would not: 
(i) The LOP mainly learns the semantic and spatial features of real objects, while other defenses focus on recognizing fake objects. (ii) The bounding boxes of cars are much larger than that of other objects, which means that there are enough samples corresponding to components of cars provided for the LOP to learn. In summary, our proposed defense incurs almost no damage on the normal performance of the victim models and may sometimes even improve the performance due to its finer granularity modeling of the obstacles. 
% In Appendix \ref{appendix:hyper}, we further experiment with the hyper-parameters of LOP, which validate that the model structure will not affect the performance of LOP, while the LOP's performance may be affected by the value of $B$ in a certain degree.
In Appendix \ref{appendix:hyper}, we further experiment with the hyper-parameters of LOP, which validate that the model structure will not affect the performance of LOP.
% The 3D object detectors with our LOP have higher precision on cars and cyclists than those detectors with other defenses, regardless of the detector structure.
% Though detectors with SRS perform better than those with our LOP on the prediction of pedestrians, SRS is considered as the baseline of these defenses in our experiment.
% Furthermore, SRS has limited influence on defending against existing appearing attacks, which will be reported in Section \ref{sec:Experiment:results:comp_attack}.
% Should analyze why LOP will improve the normal performance of a model

% Similarly, we evaluate the performance of these detectors under both normal circumstance and attacks. 

\subsubsection{Overhead Analysis}
\label{sec:Experiment:results:overhead}
Next, we evaluate the overhead in the preparation stage. Except for our LOP and SVF, other defense methods do not introduce additional learning modules and therefore no training is required in the preparation stage. Table \ref{tab:EXP4-1} compares the time overhead of LOP and SVF during the preparation phase. Table \ref{tab:EXP4-2} reports the time and the space overhead of the inference phase of each defense. We conduct the experiments with $5$ repetitive tests on each case, and report the mean and the standard deviation as the final results. 

\begin{table}[ht]
  \centering
  % \vspace{-0.2in}
  \caption{The time overhead of LOP and SVF during the preparation phase. \scriptsize{``*'' means that the results are from the OpenPCDet, an open source platform of 3D object detectors, which we use the training time of the specific 3D object detector to approximate the re-training time of SVF on the same detector.}}
\scalebox{0.7}{
  \begin{tabular}{cccc}
    \toprule
     Defense &  Total Time (h)  & Time Per Epoch (s)  & Time Per Iter (s)  \\
    \midrule
    SVF (PointPillar) & $1.2^*$  & $54.0^*$ & $8.21^*$ \\
    SVF (PointRCNN) & $3.0^*$  & $135.0^*$ & $20.51^*$ \\
    SVF (PV-RCNN) & $5.0^*$  & $225.0^*$ & $34.19^*$ \\
     Ours (PointNet) & 0.41  & 7.30  & 0.07 \\
     Ours (DGCNN) & 0.77  & 13.88 & 0.14 \\
    \bottomrule
    \end{tabular}%
}
  \label{tab:EXP4-1}%
\end{table}%

\begin{table}[ht]
  \centering
  % \vspace{-0.2in}
  \caption{The time and space overhead of LOP and other defenses during the inference phase.}
\scalebox{0.7}{
    \begin{tabular}{cccc}
    \toprule
          &  Time per sample (s)  &  GPU Mem (MB)  &  CPU Mem (MB)  \\
    \midrule
     None &  0.060±0.005 & 1477  & 2551 \\
    \midrule
     SRS  &  0.069±0.007 & 1473  & 2549 \\
     SOR  &  0.114±0.005 & 5827  & 2516 \\
     % SOR-liu &  0.111±0.004 & 5827  & 2514 \\
    \midrule
     Carlo (LPD)  &  0.503±0.003 & 1477  & 2552 \\
     Carlo (FSD)  &  2.463±0.005 & 1477  & 2506 \\
     Shadow-Catcher &  0.089±0.002 & 1477  & 2551 \\
    \midrule
     Ours (PointNet) &  1.341±0.011 & 2283  & 2518 \\
     Ours (DGCNN) &  1.589±0.013 & 3747  & 2506 \\
    \bottomrule
    \end{tabular}%
}
  \label{tab:EXP4-2}%
\end{table}%

\noindent$\bullet$\textbf{ Results \& Analysis.} As Table \ref{tab:EXP4-1} shows, the time overhead of SVF in the preparation phase is much more higher than that of LOP. It is mainly because SVF requires the retraining of the whole 3D object detectors from scratch, while the training task of LOP only involves a PC-based binary classifier, a much easier learning task compared with that of SVF. More importantly, once LOP is trained, it can be combined with different defense targets to provide the defense, while SVF has to retrain each target.

Meanwhile, Table \ref{tab:EXP4-2} shows, LOP incurs slightly more time and space overheads than most of the statistical defenses, which 
can be further reduced by 
some optimization techniques. For example, to simplify the implementation, we split the whole input space into pillars and use LOP to predict their objectness score during the split in this experiment. However, there is not necessary to check all pillars in the real case. We can identify the pillars which not only intersects with the predicted bounding boxes but also contains points, and only predict their objectness scores to reduce the total times of calculations. Furthermore, we can combine parts of these pillars into a batch and uses LOP to predict in a parallel way for further acceleration.
% a more parallel implementation of our defense. For example, the prediction over different pillars can be paralleled in our defense. 
In Section \ref{sec:Experiment:results:real}, we 
% further optimize the implementation in
follow the optimization mentioned above to deploy LOP in
the end-to-end self-driving system and reduce the time overhead caused by LOP to less than $10$ms per detection, which has almost no influence on the real-time requirement of the self-driving system. 

% In summary, compared with the heavy-weight defense SVF, our LOP has much lower time overhead during the preparation phase, while, compared with the post-processing defenses, LOP only has a slightly higher time and space overhead in both modular and end-to-end testing, which, from our perspective, is an acceptable trade-off considering the advantages in robustness enhancement.

\subsection{Adaptive Attacks}
\label{sec:Experiment:results:adaptive}
In this part, we evaluate whether our defense is robust against an adaptive attacker who knows the existence of LOP and in the worst case has the access to the structure and the parameters of our LOP. In this almost worst-case threat model, it is possible for the adversary to attempt to bypass our defense during the generation of forged objects. As the \textit{Physical} attack in \cite{jiachen2020towards} requires no training stage in its generation, we choose to modify the \textit{Baseline} attack, i.e., the attack in \cite{yulong2019advSensor}, which we refer to as the \textit{Baseline} attack throughout this response letter, into an adaptive attack against our defense. Specifically, we propose to generate the adversarial point cloud by simultaneously optimizing the original appearing attack objective and maximizing the score of the crafted object under LOP. To enhance the performance of the Baseline attack, we further replace the FGSM algorithm by PGD. Table \ref{tab:EXP1} reports the ASR of the adaptive attacks on the three 3D object detectors when LOP is deployed or not, along with the AP and the precision of the detectors on cars under the adaptive attack. 

\begin{table*}[ht]
  \centering
  % \vspace{-0.2in}
  \caption{The performance of LOP against adaptive attack. The names behind ``w/o defense'' denotes the target LOP of attack.}
\scalebox{0.9}{
    \begin{tabular}{cccccccccc}
    \toprule
          & \multicolumn{3}{c}{PointPillars} & \multicolumn{3}{c}{PointRCNN} & \multicolumn{3}{c}{PV-RCNN} \\
    \cmidrule(lr){2-4} \cmidrule(lr){5-7} \cmidrule(lr){8-10} 
          & ASR   & AP    & Precision & ASR   & AP    & Precision & ASR   & AP    & Precision \\
    \midrule
    No Attack & /     & 72.34\% & 78.99\% & /     & 75.13\% & 75.04\% & /     & 73.32\% & 73.12\% \\
    \midrule
    w/o. defense (PointNet) & 45.61\% & 68.52\% & 66.54\% & 8.78\% & 66.14\% & 55.63\% & 12.58\% & 66.37\% & 28.12\% \\
    w/o. defense (DGCNN) & 44.78\% & 68.53\% & 66.65\% & 8.56\% & 65.87\% & 55.13\% & 12.53\% & 66.39\% & 28.18\% \\
    \midrule
    Ours (PointNet, B=0.5) & 22.17\% & 69.36\% & 74.11\% & 5.14\% & 71.95\% & 74.31\% & 6.42\% & 69.27\% & 60.29\% \\
    Ours (PointNet, B=0.6) & 16.17\% & 69.55\% & 75.87\% & 4.42\% & 72.33\% & 76.06\% & \textbf{5.58\%} & 66.39\% & 61.20\% \\
    Ours (DGCNN, B=0.5) & 17.11\% & \textbf{69.73\%} & 76.30\% & 3.86\% & 72.31\% & 76.41\% & 6.58\% & \textbf{69.89\%} & 63.62\% \\
    Ours (DGCNN, B=0.6) & \textbf{12.53\%} & 68.56\% & \textbf{78.02\%} & \textbf{3.25\%} & \textbf{73.33\%} & \textbf{77.85\%} & 5.83\% & 68.37\% & \textbf{64.48\%} \\
    \bottomrule
    \end{tabular}%
}
  \label{tab:EXP1}%
\end{table*}%

\noindent$\bullet$\textbf{ Results \& Analysis.} As we can see from Table \ref{tab:EXP1}, our LOP performs well when defending against the adaptive attacks above. Both the PointNet-based and the DGCNN-based LOP can reduce the ASR of the adaptive attacks by a large margin, while only a slight loss of performance on clean samples is observed. For example, when defending PointPillars, the ASR is reduced from $45\%$ to $12\%$ with the DGCNN-based LOP, while the decrease of AP is by less than $4\%$. From our perspective, the result may be because the orthogonality between the original attack target and the intention of bypassing LOP, which brings challenges for optimizing two different loss function at the same time.  In summary, LOP has certain robustness against even the worst-case adaptive attack where the attack has a full white-box access to the defense module.

\subsection{System Integration}
\label{sec:Experiment:results:real}
 To evaluate the system-level usefulness of our proposed defense, we implement the PointNet-based LOP in Baidu's Apollo 6.0.0 system
 \xqf{in the optimized way described in Section \ref{sec:Experiment:results:overhead}}
 , and conduct both the modular and the closed-loop control evaluation in two simulation environments in normal driving scenarios and against the \textit{Physical} attack. We release the implementation details
 % and the code 
 in \cite{Apollo_LOP}.

\begin{figure*}[t]
    \centering
    \includegraphics[width=0.9\textwidth]{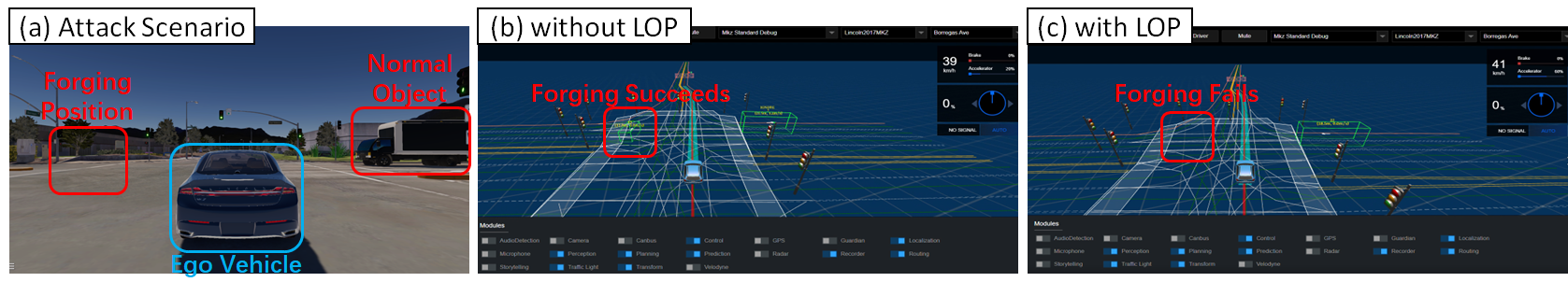}
    \vspace{-0.15in}
    \caption{The simulating scenario and the Dreamview of Apollo 6.0.0 without and with our LOP under Physical attack.}
    \label{fig:apollo_LOP}
\end{figure*}

\noindent$\bullet$\textbf{ Experimental Settings.} In the experiments, we construct two different scenarios (e.g., Single Lane Road and Borregas Ave) with random traffic in the LGSVL simulator to evaluate LOP's performance in the end-to-end system. 
% As the modular testing results, 
Table \ref{tab:EXP6} reports the ASR of the Physical attack on Apollo 6.0.0, together with the precision and the time cost of the 3D object detectors in Apollo's perception module when LOP is deployed or not, and 
% As the closed-loop control testing results,
Fig.\ref{fig:apollo_LOP} illustrates the detection results in an end-to-end driving test 
% (i.e., closed-loop control with Perception, Prediction, Localization and Planning modules enabled) 
when the system is deployed without or with LOP, and shows a snapshot of the attacking scenario
% in the \textit{Borregas Ave} map 
in the experiments.

\begin{table}[ht]
  \centering
  % \vspace{-0.2in}
  % \caption{The performance of the perception module in the Apollo system deployed with or without LOP.}
  \caption{The performance of the perception module in an end-to-end Apollo 6.0.0 system when the 3D object detector is deployed with or without LOP.}
\scalebox{0.75}{
    \begin{tabular}{lcccc}
    \toprule
          & Precision & ASR   & time cost (ms) & FPS \\
    \midrule
    Apollo 6.0.0 (w/o. LOP) & 8.33\% & 53.66\% & 33.36ms & 29.97 \\
    Apollo 6.0.0 (w/. LOP) & 100.00\% & 0.00\% & 42.48ms & 23.54 \\
    \bottomrule
    \end{tabular}%
}
  \label{tab:EXP6}%
\end{table}%

%%%%%%%% BEGIN APOLLO
%%%%%%%% END APOLLO

\noindent$\bullet$\textbf{ Results \& Analysis.} As Table \ref{tab:EXP6} shows, LOP effectively defends against appearing attacks in the end-to-end Apollo 6.0.0, with a slight proportion of time overhead (less than $10$ms). As Fig.\ref{fig:apollo_LOP}(b) shows, the \textit{Physical} attack can successfully fools Apollo's perception module, and remains existent in the Dreamview even after the processing of MOT. This confirms our argument that appearing attacks is easier to be mounted in practical scenarios than disappearing attacks. In the Dreamview view of Fig.\ref{fig:apollo_LOP}(c), with the help of our LOP, the forged object is eliminated from Apollo's perception during the evaluation (with ASR$=0\%$), while the real obstacles remain intact in the perception of the ADS. Therefore, the driving trajectory of the ADS with LOP remains normal and safe during the full driving test. Besides, LOP only incurs a $9.12ms$ overhead on the running time of the 3D detection pipeline on average and slightly brings down the FPS from $29.97$ to $23.54$, which still satisfies the real-time requirement of a physical self-driving system \cite{yulan2021deep}.  

% Moreover, we select previous forged objects, which can successfully fool the perception module of Apollo for at least one frame, to further determine whether they would lead to a potential car crash in different traces by showing the crash rate. The crash rate is calculated based on the assumption that there is a car moving behind the self-driving vehicle at a constant speed, which means some specific events of the Apollo' planning may lead to the crash between the imaginary car and the self-driving vehicle, such as the sudden brake. We observe that the crash rate of the Apollo without LOP is $13.33\%$ $(2/15)$, while, with LOP, the crash rate is reduced to $0.00\%$ $(0/15)$. We provide the Dreamview snapshot and the details of these experiments in \xqf{Appendix \ref{appendix:system}}. Therefore, combined with the comprehensive evaluation results on the KITTI benchmark, our end-to-end experiments further validate the system-level usefulness of our proposed defense in terms of the improved system robustness, and the acceptable overhead on the running time and the normal driving performance.

Moreover, we use the previously forged objects, which can successfully fool the perception module of Apollo for at least one frame, to further test whether they would lead to a potential harsh braking in different traces. Specifically, we measure whether the self-driving vehicle would do sudden braking, 
which is shown as it decelerating to 0 km/h in less than 1 second, 
to calculate the \textit{harsh braking rate}, i.e., the ratio of the test cases where the self-driving vehicle suddenly brakes when there is no real obstacle in front of it. We observe that the harsh braking rate of the Apollo without LOP is $13.33\%$ $(2/15)$, while, with LOP, the harsh braking rate is reduced to $0.00\%$ $(0/15)$. We provide the Dreamview snapshot and the details of these experiments in \xqf{Appendix \ref{appendix:system}}. Therefore, combined with the comprehensive evaluation results on the KITTI benchmark, our end-to-end experiments further validate the system-level usefulness of our proposed defense in terms of the improved system robustness, and the acceptable overhead on the running time and the normal driving performance.

% Furthermore, we also evaluate the effectiveness of our defense with an experimental vehicle for automated driving, the D-KIT Advanced \cite{DKit}, where real-world PCs are captured by the Velodyne 128 LiDAR of it when the ADS drives in a closed environment.
% with 5 moving obstacles. 
% We report and analyze the experimental results in Appendix \ref{app:realworld}, which validates the practical usefulness of our defense from a different aspect.  

% \input{tex/ablation}
\section{Discussion}\label{sec:Discussion}

% Our current defense mainly focuses on appearing attacks, which form a popular attack class on LiDAR-based object detectors in ADS. Below, we discuss other attack classes which may threaten the detector's security and why our current work chooses the appearing attack as the main research target.

\noindent\textbf{Appearing Attacks vs. Disappearing Attacks.} Our current defense mainly focuses on appearing attacks, which form a popular attack class on LiDAR-based object detectors in ADS
In contrast to appearing attacks, a disappearing attack aims at hiding the existing objects from the prediction results of the victim 3D object detector \cite{dawn2018physical,shang2018shapeshifter,yue2019seeing}. To accomplish this purpose, the adversary would optimally generate a 3D-printing object to the target detector would not recognize it or its neighbouring object, and put the object on the road or near some objects to mount the attack. 
% Technically, the specific shape of the printed object is optimally chosen to guarantee the target detector would not recognize it or its neighbouring object in the prediction results, leading to its disappearance in the detector's output.  

In the previous literature, Cao et al. propose one of the earliest disappearing attacks on ADS, and successfully hide the printed objects from the LiDAR-based detection system of Baidu's Apollo by modeling its preprocessing and postprocessing phases into differentiable functions \cite{yulong2019advObj}.
Later, Tu et al. present a more general disappearing attack which breaks the state-of-the-art 3D object detectors including PointPillars and PointRCNN, and hide the car on which the printed object is positioned from the model's prediction results \cite{james2020physically}.
Recently, Cao et al. further devise a more powerful disappearing attack, MSF-ADV, which fools the image-based 2D object detectors and LiDAR-based 3D object detectors at the same time, and causes the fusion-based detection system of Baidu's Apollo to ignore the existence of the printed objects \cite{yulong2021invisible}.

Compared with appearing attack, we argue that a disappearing attack is not physical because it is untargeted and \textit{single-shot}, i.e., the attacker has to put a printed object on the road or near some objects in preparation. This indicates that he/she could hardly choose the victim ADS during the attack. Moreover, the printed object can only take effect once because it might be destroyed or recognized by the people nearby after the first accident happens. In contrast, in an appearing attack the attacker can choose the victim to fire the laser and forge non-existent cars as he/she wishes, making it difficult for others to note the attack due to the almost no evidence left in the accident scene. Nevertheless, considering the severe consequences if happening, how to mitigate disappearing attacks remains a meaningful direction to pursue.
% for future works.

\noindent\textbf{Extension to Other Attack Classes.} We further discuss the applicability of our defense for mitigating mis-categorization attacks, 
% which aims at changing the predicted class of the target objects in the prediction results of the victim 3D object detector. 
which aims at changing the predicted class of the target objects in the victim's detection results. 
In this sense, a mis-categorization attack can be seen as the combination of a disappearing attack and an appearing attack. In the above process, we observe that the crafted object would also be left with an abnormal density-depth characteristic which does not belong to the target class. 
% Therefore we modify the appearing attacks covered in our experiments into mis-categorization attacks, which selects the objects from the \textit{bicycle} or \textit{pedestrian} classes, and injecting a limited number of points around them to fool the victim 3D detector to mis-categorize them as vehicles, and evaluate the performance of our LOP when deployed alongside the 3D detector. The experimental results prove that our proposed defense is also effective against mis-categorization attacks due to the depth-density anomaly introduced by them.
Specifically, in Appendix \ref{appendix:miscategorization}, we modify the appearing attacks covered in our experiments into mis-categorization attacks, which selects the objects from the \textit{bicycle} or \textit{pedestrian} classes, and injecting a limited number of points around them to fool the victim 3D detector to mis-categorize them as vehicles,  and evaluate the performance of our LOP when deployed alongside the 3D detector. 
% The experimental results in Appendix \ref{appendix:miscategorization} show that our proposed defense is also effective against mis-categorization attacks. It is probably due to the depth-density anomaly introduced by the mis-categorization attack.
The experimental results in Appendix \ref{appendix:miscategorization} show that our proposed defense is also effective against mis-categorization attacks due to the depth-density anomaly introduced by them.

\noindent\textbf{Fusion Models as Defense Targets.}  We first clarify the relation between our proposed LOP and the fusion models. According to \cite{yulan2021deep}, the detection frequency of existing fusion models (including FPN, FCN and AVOD) is usually lower than $15$ FPS, and may be unsuitable for real-time self-driving systems due to the efficiency bottleneck. Besides, we suggest our defense is orthogonal to the fusion strategy. LOP in our defense provides a different view for the detectors to confirm their detection, while the fusion strategy incorporates new input modality to enhance robustness. Therefore, instead of viewing fusion models as a comparison group to our defense, we prefer to view the fusion models, which are by essence detectors, as our defense targets. In Appendix \ref{appendix:fusion}, we provide a preliminary study which validates that our LOP substantially improves the robustness of fusion models against appearing attacks. For example, the PointNet-based LOP would reduce the ASR of the \textit{Physical} attack on EPNet \cite{Huang2020EPNetEP} to $0\%$. In other words, we prefer not to view LOP as a competitor for the fusion models. Instead, LOP empirically improves the robustness of the fusion models, while, as no modifications is made on the image input branch, LOP would not hurt the benefits of fusion models in self-driving systems. For future works, it would be meaningful to systematically evaluate our proposed defense on more representative fusion and 3D object detection models. 

% Yet, we think that our LOP does not conflict with the fusion strategy, and the fusion model should be seen as a defense target as well as other 3D object detectors.

\noindent\textbf{Limitation and Future Directions.}  Finally, we discuss the potential limitations of our proposed defense:
% In Appendix \ref{appendix:FP}, we present a further case study on these false positives from our defense. We find the number of false positives with depth less than $10$ meters is only $1.60\%$ of the total real vehicles, and the number of false positives with depth less than $20$ meters is only $2.21\%$ of the total real vehicles. The results imply that our LOP may not recognize the forged obstacles well in some cases due to its uncertainty on distant vehicles.
According to the case study on the false positives from our defense, we find that our LOP may not recognize the forged obstacles well in some cases due to its uncertainty on distant vehicles. However, due to the existence of the MOT module, the self-driving system keeps refreshing the driving plan and corrects the mis-prediction of distant objects when the obstacle comes nearby. Moreover, MOT would prevent the self-driving system from ignoring a distant object only if LOP misses a distant object in several consecutive frames, the possibility of which is less than $0.1\%$ according to our calculation. Therefore, the negative influence of LOP on the normal performance of the detector would hardly influence the normal driving behaviors of the defense target. The similar results are also provided in our end-to-end experiments in Section \ref{sec:Experiment:results:real}.

Besides, due to our limited computing resources, we mainly prove the advantages of LOP in terms of computational overhead compared with SVF, while we admit the additional overhead may trade for better defense effectiveness and would not be a problem for most autonomous driving companies. Nevertheless, SVF as a retraining-based approach lies in a different defense category from our proposed plug-and-play defense module. A 3D object detection module which is enhanced by SVF can be further combined with our LOP for better defense effectiveness. As SVF still has a space for improvement in defense effectiveness \cite{jiachen2020towards}, it would be meaningful for future works to explore their combination in the future.

\section{Conclusion} \label{sec:Conclusion}
In this paper, we systematically analyze the working mechanisms of recent appearing attacks and summarize their common weaknesses in violating the depth-density law and failing to imitate the local parts of real objects.
Based on the defensive insights, we propose a novel plug-and-play defense method which adopts a LOP module to work by side of an arbitrary LiDAR-based object detector to detect and eliminate forged obstacles from its prediction results.
To handle the complexity of the depth-density law and the local object feature, we build the LOP with an off-the-shelf point-wise PC classifier and explicitly expand the input point feature with the derived depth information.
We present extensive experiments spanning three state-of-the-art 3D object detectors and three known appearing attacks
% , on both the standard benchmark KITTI dataset and real-world PC data we collect from a real-car test, 
on the standard benchmark KITTI dataset,
which validate the effectiveness and flexibility of our proposed defense.
Furthermore, we deploy and evaluate the LOP in an end-to-end self-driving system, which validates the system-level usefulness of our proposed defense.
% To validate the effectiveness and flexibility of our proposed defense, we conduct extensive experiments covering three state-of-the-art 3D object detectors and three known appearing attacks, with both the standard benchmark KITTI and real-world PC data we collect from a real-car test.

%% 说我们的方法时不能用spatial和semantic

% \louis{Experiment results show} that our LOP can improve the performance of three state-of-the-art 3D object detectors in normal circumstance or under attacks.
% We further gather the PC data sampled by LiDAR \louis{in a closed road environment}, and prove that our LOP can also perform well in the real world.

% We conjecture that to better defend against appearing attacks or even other latent adversarial attacks, it is necessary to fuse multiple object detectors based-on \louis{different sources of input}, and deploy related defense on each object detector participated in this fusion.
% Thus, our work offers the basis for designing such robust object detectors for ADS.
\section*{Acknowledgments}
We would like to thank the anonymous reviewers and the shepherd for their insightful comments that helped improve the quality of the paper. This work was supported in part by the National Key Research and Development Program (2021YFB3101200), National Natural Science Foundation of China (61972099, U1736208, U1836210, U1836213, 62172104, 62172105, 61902374, 62102093, 62102091). Min Yang is a faculty of Shanghai Institute of Intelligent Electronics \& Systems, Shanghai Institute for Advanced Communication and Data Science, and Engineering Research Center of Cyber Security Auditing and Monitoring, Ministry of Education, China. Mi Zhang and Min Yang are the corresponding authors.

% This work was supported in part by National Natural Science Foundation of China (61972099, U1836213,U1836210, U1736208), and Natural Science Foundation of Shanghai (19ZR1404800). Min Yang is a faculty of Shanghai Institute of Intelligent Electronics \& Systems, Shanghai Institute for Advanced Communication and Data Science, and Engineering Research Center of CyberSecurity Auditing and Monitoring, Ministry of Education, China. 

% \clearpage

\bibliographystyle{plain}
\bibliography{ref}

\begin{thebibliography}{10}

\bibitem{Apollo}
{Apollo Open Platform}.
\newblock \url{https://apollo.auto/developer.html}.
\newblock Accessed: 2022-01-30.

\bibitem{apollo_code}
{ApolloAuto/Apollo}.
\newblock \url{https://github.com/ApolloAuto/apollo/tree/master}.

\bibitem{Apollo_LOP}
{Apollo\_LOP}.
\newblock \url{https://anonymous.4open.science/r/Apollo_LOP-A1F4}.

\bibitem{DKit}
{Baidu Autonomous Driving Development Kit (Apollo D-KIT)}.
\newblock \url{https://apollo.auto/apollo\_d\_kit.html}.
\newblock Accessed: 2022-01-30.

\bibitem{waymo_report}
{Combine Lidar and Cameras for 3D object detection - Waymo}.
\newblock \url{https://www.louisbouchard.ai/waymo-lidar/}.

\bibitem{Waymo}
{Waymo One - Waymo}.
\newblock \url{https://waymo.com/waymo-one/}.
\newblock Accessed: 2022-01-30.

\bibitem{LiDAR}
Light detection and ranging.
\newblock In Shashi Shekhar, Hui Xiong, and Xun Zhou, editors, {\em
  Encyclopedia of {GIS}}, page 1119. Springer, 2017.

\bibitem{marco2016MLinTrack}
Marco Allodi, Alberto Broggi, Domenico Giaquinto, Marco Patander, and Antonio
  Prioletti.
\newblock Machine learning in tracking associations with stereo vision and
  lidar observations for an autonomous vehicle.
\newblock In {\em 2016 {IEEE} Intelligent Vehicles Symposium, {IV} 2016,
  Gotenburg, Sweden, June 19-22, 2016}, pages 648--653. {IEEE}, 2016.

\bibitem{pierre2019lunet}
Pierre Biasutti, Vincent Lepetit, Jean{-}Fran{\c{c}}ois Aujol, Mathieu
  Br{\'{e}}dif, and Aur{\'{e}}lie Bugeau.
\newblock Lu-net: An efficient network for 3d lidar point cloud semantic
  segmentation based on end-to-end-learned 3d features and u-net.
\newblock In {\em 2019 {IEEE/CVF} International Conference on Computer Vision
  Workshops, {ICCV} Workshops 2019, Seoul, Korea (South), October 27-28, 2019},
  pages 942--950. {IEEE}, 2019.

\bibitem{yulong2021invisible}
Yulong Cao, Ningfei Wang, Chaowei Xiao, Dawei Yang, Jin Fang, Ruigang Yang,
  Qi~Alfred Chen, Mingyan Liu, and Bo~Li.
\newblock Invisible for both camera and lidar: Security of multi-sensor fusion
  based perception in autonomous driving under physical-world attacks.
\newblock In {\em 42nd {IEEE} Symposium on Security and Privacy, {SP} 2021, San
  Francisco, CA, USA, 24-27 May 2021}, pages 176--194. {IEEE}, 2021.

\bibitem{yulong2019advSensor}
Yulong Cao, Chaowei Xiao, Benjamin Cyr, Yimeng Zhou, Won Park, Sara Rampazzi,
  Qi~Alfred Chen, Kevin Fu, and Z.~Morley Mao.
\newblock Adversarial sensor attack on lidar-based perception in autonomous
  driving.
\newblock In {\em Proceedings of the 2019 {ACM} {SIGSAC} Conference on Computer
  and Communications Security, {CCS} 2019, London, UK, November 11-15, 2019},
  pages 2267--2281. {ACM}, 2019.

\bibitem{yulong2019advObj}
Yulong Cao, Chaowei Xiao, Dawei Yang, Jing Fang, Ruigang Yang, Mingyan Liu, and
  Bo~Li.
\newblock Adversarial objects against lidar-based autonomous driving systems.
\newblock {\em CoRR}, abs/1907.05418, 2019.

\bibitem{nicholas2017towards}
Nicholas Carlini and David~A. Wagner.
\newblock Towards evaluating the robustness of neural networks.
\newblock In {\em 2017 {IEEE} Symposium on Security and Privacy, {SP} 2017, San
  Jose, CA, USA, May 22-26, 2017}, pages 39--57. {IEEE} Computer Society, 2017.

\bibitem{depth_density}
Jamie Carter, Keil Schmid, Kirk Waters, Lindy Betzhold, Brian Hadley, Rebecca
  Mataosky, and Jennifer Halleran.
\newblock Lidar 101: An introduction to lidar technology, data, and
  applications.
\newblock {\em National Oceanic and Atmospheric Administration (NOAA) Coastal
  Services Center}, 2012.

\bibitem{qi2020object}
Qi~Chen, Lin Sun, Zhixin Wang, Kui Jia, and Alan~L. Yuille.
\newblock Object as hotspots: An anchor-free 3d object detection approach via
  firing of hotspots.
\newblock In {\em Computer Vision - {ECCV} 2020 - 16th European Conference,
  Glasgow, UK, August 23-28, 2020, Proceedings, Part {XXI}}, volume 12366,
  pages 68--84. Springer, 2020.

\bibitem{shang2018shapeshifter}
Shang{-}Tse Chen, Cory Cornelius, Jason Martin, and Duen Horng~(Polo) Chau.
\newblock Shapeshifter: Robust physical adversarial attack on faster {R-CNN}
  object detector.
\newblock In {\em Machine Learning and Knowledge Discovery in Databases -
  European Conference, {ECML} {PKDD} 2018, Dublin, Ireland, September 10-14,
  2018, Proceedings, Part {I}}, volume 11051 of {\em Lecture Notes in Computer
  Science}, pages 52--68. Springer, 2018.

\bibitem{hsu2021P3DMOT}
Hsu{-}Kuang Chiu, Jie Li, Rares Ambrus, and Jeannette Bohg.
\newblock Probabilistic 3d multi-modal, multi-object tracking for autonomous
  driving.
\newblock In {\em {IEEE} International Conference on Robotics and Automation,
  {ICRA} 2021, Xi'an, China, May 30 - June 5, 2021}, pages 14227--14233.
  {IEEE}, 2021.

\bibitem{jacob2019bert}
Jacob Devlin, Ming{-}Wei Chang, Kenton Lee, and Kristina Toutanova.
\newblock {BERT:} pre-training of deep bidirectional transformers for language
  understanding.
\newblock In {\em Proceedings of the 2019 Conference of the North American
  Chapter of the Association for Computational Linguistics: Human Language
  Technologies, {NAACL-HLT} 2019, Minneapolis, MN, USA, June 2-7, 2019, Volume
  1 (Long and Short Papers)}, pages 4171--4186. Association for Computational
  Linguistics, 2019.

\bibitem{andreas2012kitti}
Andreas Geiger, Philip Lenz, and Raquel Urtasun.
\newblock Are we ready for autonomous driving? the {KITTI} vision benchmark
  suite.
\newblock In {\em 2012 {IEEE} Conference on Computer Vision and Pattern
  Recognition, Providence, RI, USA, June 16-21, 2012}, pages 3354--3361. {IEEE}
  Computer Society, 2012.

\bibitem{ian2015explaining}
Ian~J. Goodfellow, Jonathon Shlens, and Christian Szegedy.
\newblock Explaining and harnessing adversarial examples.
\newblock In {\em 3rd International Conference on Learning Representations,
  {ICLR} 2015, San Diego, CA, USA, May 7-9, 2015, Conference Track
  Proceedings}, 2015.

\bibitem{yulan2021deep}
Yulan Guo, Hanyun Wang, Qingyong Hu, Hao Liu, Li~Liu, and Mohammed Bennamoun.
\newblock Deep learning for 3d point clouds: {A} survey.
\newblock {\em {IEEE} Trans. Pattern Anal. Mach. Intell.}, 43(12):4338--4364,
  2021.

\bibitem{Hallyburton2021SecurityAO}
R~Spencer Hallyburton, Yupei Liu, Yulong Cao, Z~Morley Mao, and Miroslav Pajic.
\newblock Security analysis of camera-lidar fusion against black-box attacks on
  autonomous vehicles.
\newblock In {\em 31st USENIX Security Symposium (USENIX SECURITY)}, 2022.

\bibitem{abdullah2020advpc}
Abdullah Hamdi, Sara Rojas, Ali~K. Thabet, and Bernard Ghanem.
\newblock Advpc: Transferable adversarial perturbations on 3d point clouds.
\newblock In {\em Computer Vision - {ECCV} 2020 - 16th European Conference,
  Glasgow, UK, August 23-28, 2020, Proceedings, Part {XII}}, volume 12357 of
  {\em Lecture Notes in Computer Science}, pages 241--257. Springer, 2020.

\bibitem{zhongyuan2021shadowcatcher}
Zhongyuan Hau, Soteris Demetriou, Luis Mu{\~{n}}oz{-}Gonz{\'{a}}lez, and
  Emil~C. Lupu.
\newblock Shadow-catcher: Looking into shadows to detect ghost objects in
  autonomous vehicle 3d sensing.
\newblock In {\em Computer Security - {ESORICS} 2021 - 26th European Symposium
  on Research in Computer Security, Darmstadt, Germany, October 4-8, 2021,
  Proceedings, Part {I}}, volume 12972 of {\em Lecture Notes in Computer
  Science}, pages 691--711. Springer, 2021.

\bibitem{Huang2020EPNetEP}
Tengteng Huang, Zhe Liu, Xiwu Chen, and Xiang Bai.
\newblock Epnet: Enhancing point features with image semantics for 3d object
  detection.
\newblock July 2020.

\bibitem{shuiwang20103DCNN}
Shuiwang Ji, Wei Xu, Ming Yang, and Kai Yu.
\newblock 3d convolutional neural networks for human action recognition.
\newblock In {\em Proceedings of the 27th International Conference on Machine
  Learning (ICML-10), June 21-24, 2010, Haifa, Israel}, pages 495--502.
  Omnipress, 2010.

\bibitem{alex2019pointpillars}
Alex~H. Lang, Sourabh Vora, Holger Caesar, Lubing Zhou, Jiong Yang, and Oscar
  Beijbom.
\newblock Pointpillars: Fast encoders for object detection from point clouds.
\newblock In {\em {IEEE} Conference on Computer Vision and Pattern Recognition,
  {CVPR} 2019, Long Beach, CA, USA, June 16-20, 2019}, pages 12697--12705.
  Computer Vision Foundation / {IEEE}, 2019.

\bibitem{yann1989backpropagation}
Yann LeCun, Bernhard~E. Boser, John~S. Denker, Donnie Henderson, Richard~E.
  Howard, Wayne~E. Hubbard, and Lawrence~D. Jackel.
\newblock Backpropagation applied to handwritten zip code recognition.
\newblock {\em Neural Comput.}, 1(4):541--551, 1989.

\bibitem{yangyan2018PCNN}
Yangyan Li, Rui Bu, Mingchao Sun, Wei Wu, Xinhan Di, and Baoquan Chen.
\newblock Pointcnn: Convolution on x-transformed points.
\newblock In {\em Advances in Neural Information Processing Systems 31: Annual
  Conference on Neural Information Processing Systems 2018, NeurIPS 2018,
  December 3-8, 2018, Montr{\'{e}}al, Canada}, pages 828--838, 2018.

\bibitem{tsungyi2017focalloss}
Tsung{-}Yi Lin, Priya Goyal, Ross~B. Girshick, Kaiming He, and Piotr
  Doll{\'{a}}r.
\newblock Focal loss for dense object detection.
\newblock In {\em {IEEE} International Conference on Computer Vision, {ICCV}
  2017, Venice, Italy, October 22-29, 2017}, pages 2999--3007. {IEEE} Computer
  Society, 2017.

\bibitem{wenhan2021motReview}
Wenhan Luo, Junliang Xing, Anton Milan, Xiaoqin Zhang, Wei Liu, and Tae{-}Kyun
  Kim.
\newblock Multiple object tracking: {A} literature review.
\newblock {\em Artif. Intell.}, 293:103448, 2021.

\bibitem{gregory2019lasernet}
Gregory~P. Meyer, Ankit Laddha, Eric Kee, Carlos Vallespi{-}Gonzalez, and
  Carl~K. Wellington.
\newblock Lasernet: An efficient probabilistic 3d object detector for
  autonomous driving.
\newblock In {\em {IEEE} Conference on Computer Vision and Pattern Recognition,
  {CVPR} 2019, Long Beach, CA, USA, June 16-20, 2019}, pages 12677--12686.
  Computer Vision Foundation / {IEEE}, 2019.

\bibitem{sina2020practical}
Sina Mohseni, Mandar Pitale, Vasu Singh, and Zhangyang Wang.
\newblock Practical solutions for machine learning safety in autonomous
  vehicles.
\newblock In {\em Proceedings of the Workshop on Artificial Intelligence
  Safety, co-located with 34th {AAAI} Conference on Artificial Intelligence,
  SafeAI@AAAI 2020, New York City, NY, USA, February 7, 2020}, volume 2560 of
  {\em {CEUR} Workshop Proceedings}, pages 162--169. CEUR-WS.org, 2020.

\bibitem{charles2017PN}
Charles~Ruizhongtai Qi, Hao Su, Kaichun Mo, and Leonidas~J. Guibas.
\newblock Pointnet: Deep learning on point sets for 3d classification and
  segmentation.
\newblock In {\em 2017 {IEEE} Conference on Computer Vision and Pattern
  Recognition, {CVPR} 2017, Honolulu, HI, USA, July 21-26, 2017}, pages 77--85.
  {IEEE} Computer Society, 2017.

\bibitem{charles2017PN++}
Charles~Ruizhongtai Qi, Li~Yi, Hao Su, and Leonidas~J. Guibas.
\newblock Pointnet++: Deep hierarchical feature learning on point sets in a
  metric space.
\newblock In {\em Advances in Neural Information Processing Systems 30: Annual
  Conference on Neural Information Processing Systems 2017, December 4-9, 2017,
  Long Beach, CA, {USA}}, pages 5099--5108, 2017.

\bibitem{yunli2021MLEnabled}
Yunli Shao, Yuan Zheng, and Zongxuan Sun.
\newblock Machine learning enabled traffic prediction for speed optimization of
  connected and autonomous electric vehicles.
\newblock In {\em 2021 American Control Conference, {ACC} 2021, New Orleans,
  LA, USA, May 25-28, 2021}, pages 172--177. {IEEE}, 2021.

\bibitem{shaoshuai2020pvrcnn}
Shaoshuai Shi, Chaoxu Guo, Li~Jiang, Zhe Wang, Jianping Shi, Xiaogang Wang, and
  Hongsheng Li.
\newblock {PV-RCNN:} point-voxel feature set abstraction for 3d object
  detection.
\newblock In {\em 2020 {IEEE/CVF} Conference on Computer Vision and Pattern
  Recognition, {CVPR} 2020, Seattle, WA, USA, June 13-19, 2020}, pages
  10526--10535. Computer Vision Foundation / {IEEE}, 2020.

\bibitem{shaoshuai2019pointrcnn}
Shaoshuai Shi, Xiaogang Wang, and Hongsheng Li.
\newblock Pointrcnn: 3d object proposal generation and detection from point
  cloud.
\newblock In {\em {IEEE} Conference on Computer Vision and Pattern Recognition,
  {CVPR} 2019, Long Beach, CA, USA, June 16-20, 2019}, pages 770--779. Computer
  Vision Foundation / {IEEE}, 2019.

\bibitem{hocheol2017illusion}
Hocheol Shin, Dohyun Kim, Yujin Kwon, and Yongdae Kim.
\newblock Illusion and dazzle: Adversarial optical channel exploits against
  lidars for automotive applications.
\newblock In {\em Cryptographic Hardware and Embedded Systems - {CHES} 2017 -
  19th International Conference, Taipei, Taiwan, September 25-28, 2017,
  Proceedings}, volume 10529 of {\em Lecture Notes in Computer Science}, pages
  445--467. Springer, 2017.

\bibitem{dawn2018physical}
Dawn Song, Kevin Eykholt, Ivan Evtimov, Earlence Fernandes, Bo~Li, Amir
  Rahmati, Florian Tram{\`{e}}r, Atul Prakash, and Tadayoshi Kohno.
\newblock Physical adversarial examples for object detectors.
\newblock In {\em 12th {USENIX} Workshop on Offensive Technologies, {WOOT}
  2018, Baltimore, MD, USA, August 13-14, 2018}. {USENIX} Association, 2018.

\bibitem{jiachen2020towards}
Jiachen Sun, Yulong Cao, Qi~Alfred Chen, and Z.~Morley Mao.
\newblock Towards robust lidar-based perception in autonomous driving: General
  black-box adversarial sensor attack and countermeasures.
\newblock In {\em 29th {USENIX} Security Symposium, {USENIX} Security 2020,
  August 12-14, 2020}, pages 877--894. {USENIX} Association, 2020.

\bibitem{Szegedy2014IntriguingPO}
Christian Szegedy, Wojciech Zaremba, Ilya Sutskever, Joan Bruna, Dumitru Erhan,
  Ian~J. Goodfellow, and Rob Fergus.
\newblock Intriguing properties of neural networks.
\newblock In {\em 2nd International Conference on Learning Representations,
  {ICLR} 2014, Banff, AB, Canada, April 14-16, 2014, Conference Track
  Proceedings}, 2014.

\bibitem{OpenPCDet}
OpenPCDet~Development Team.
\newblock {OpenPCDet: An Open-source Toolbox for 3D Object Detection from Point
  Clouds}.
\newblock \url{https://github.com/open-mmlab/OpenPCDet}, 2020.

\bibitem{james2020physically}
James Tu, Mengye Ren, Sivabalan Manivasagam, Ming Liang, Bin Yang, Richard Du,
  Frank Cheng, and Raquel Urtasun.
\newblock Physically realizable adversarial examples for lidar object
  detection.
\newblock In {\em 2020 {IEEE/CVF} Conference on Computer Vision and Pattern
  Recognition, {CVPR} 2020, Seattle, WA, USA, June 13-19, 2020}, pages
  13713--13722. Computer Vision Foundation / {IEEE}, 2020.

\bibitem{yue2019DGCNN}
Yue Wang, Yongbin Sun, Ziwei Liu, Sanjay~E. Sarma, Michael~M. Bronstein, and
  Justin~M. Solomon.
\newblock Dynamic graph {CNN} for learning on point clouds.
\newblock {\em {ACM} Trans. Graph.}, 38(5):146:1--146:12, 2019.

\bibitem{xinshuo2020AB3DMOT}
Xinshuo Weng, Jianren Wang, David Held, and Kris Kitani.
\newblock 3d multi-object tracking: {A} baseline and new evaluation metrics.
\newblock In {\em {IEEE/RSJ} International Conference on Intelligent Robots and
  Systems, {IROS} 2020, Las Vegas, NV, USA, October 24, 2020 - January 24,
  2021}, pages 10359--10366. {IEEE}, 2020.

\bibitem{chong2019generating}
Chong Xiang, Charles~R. Qi, and Bo~Li.
\newblock Generating 3d adversarial point clouds.
\newblock In {\em {IEEE} Conference on Computer Vision and Pattern Recognition,
  {CVPR} 2019, Long Beach, CA, USA, June 16-20, 2019}, pages 9136--9144.
  Computer Vision Foundation / {IEEE}, 2019.

\bibitem{kaichen2021robust}
Kaichen Yang, Tzungyu Tsai, Honggang Yu, Max Panoff, Tsung{-}Yi Ho, and Yier
  Jin.
\newblock Robust roadside physical adversarial attack against deep learning in
  lidar perception modules.
\newblock In {\em {ASIA} {CCS} '21: {ACM} Asia Conference on Computer and
  Communications Security, Virtual Event, Hong Kong, June 7-11, 2021}, pages
  349--362. {ACM}, 2021.

\bibitem{yue2020onisomety}
Yue Zhao, Yuwei Wu, Caihua Chen, and Andrew Lim.
\newblock On isometry robustness of deep 3d point cloud models under
  adversarial attacks.
\newblock In {\em 2020 {IEEE/CVF} Conference on Computer Vision and Pattern
  Recognition, {CVPR} 2020, Seattle, WA, USA, June 13-19, 2020}, pages
  1198--1207. Computer Vision Foundation / {IEEE}, 2020.

\bibitem{yue2019seeing}
Yue Zhao, Hong Zhu, Ruigang Liang, Qintao Shen, Shengzhi Zhang, and Kai Chen.
\newblock Seeing isn't believing: Towards more robust adversarial attack
  against real world object detectors.
\newblock In {\em Proceedings of the 2019 {ACM} {SIGSAC} Conference on Computer
  and Communications Security, {CCS} 2019, London, UK, November 11-15, 2019},
  pages 1989--2004. {ACM}, 2019.

\bibitem{hang2019dupnet}
Hang Zhou, Kejiang Chen, Weiming Zhang, Han Fang, Wenbo Zhou, and Nenghai Yu.
\newblock Dup-net: Denoiser and upsampler network for 3d adversarial point
  clouds defense.
\newblock In {\em 2019 {IEEE/CVF} International Conference on Computer Vision,
  {ICCV} 2019, Seoul, Korea (South), October 27 - November 2, 2019}, pages
  1961--1970. {IEEE}, 2019.

\bibitem{yin2018VoxelNet}
Yin Zhou and Oncel Tuzel.
\newblock Voxelnet: End-to-end learning for point cloud based 3d object
  detection.
\newblock In {\em 2018 {IEEE} Conference on Computer Vision and Pattern
  Recognition, {CVPR} 2018, Salt Lake City, UT, USA, June 18-22, 2018}, pages
  4490--4499. Computer Vision Foundation / {IEEE} Computer Society, 2018.

\bibitem{ji2018onlineMOT}
Ji~Zhu, Hua Yang, Nian Liu, Minyoung Kim, Wenjun Zhang, and Ming{-}Hsuan Yang.
\newblock Online multi-object tracking with dual matching attention networks.
\newblock In {\em Computer Vision - {ECCV} 2018 - 15th European Conference,
  Munich, Germany, September 8-14, 2018, Proceedings, Part {V}}, volume 11209
  of {\em Lecture Notes in Computer Science}, pages 379--396. Springer, 2018.

\end{thebibliography}

\appendix
\section{Hyperparameter Sensitivity}
\label{appendix:hyper}
\begin{figure*}[t]
    \centering
    \includegraphics[width=0.9\textwidth]{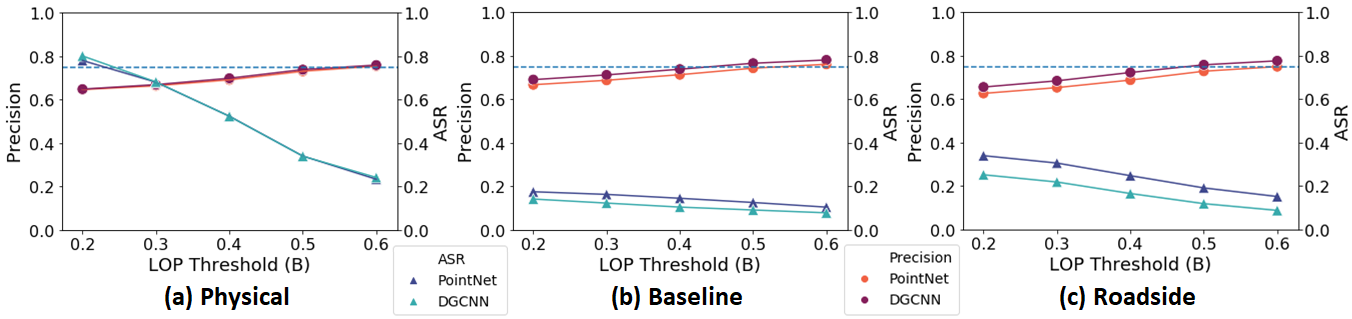}
    \caption{The ASR of different attacks and the precision of PointRCNN when deploying the LOP with different values of $B$ and different model structures on PointRCNN.
    % \lyf{The precision and ASR metrics are represented in triangles and circles respectively, while the structures of LOP are distinguished in colors.}
    % The blue horizontal dotted lines in (a), (b) and (c) all represent the precision of PointRCNN on cars in the normal circumstances.
    }
    \label{fig:ablation_atk}
\end{figure*}

\noindent\textbf{Experiment Settings.} To explore the impact of the hyperparameters and the structure of LOP, we implement the LOP with different values of $B$ and different architectures. 
Table \ref{tab:ablation_normal} report the AP and precision of PointRCNN when our defense is deployed with different settings under the normal circumstance. Fig.\ref{fig:ablation_atk} report the defense effectiveness and the precision of PointRCNN when our defense is deployed with different settings under attacks.

%%%%%%% BEGIN Ablation

\begin{table}[ht]
  \centering
  \vspace{-0.2in}
  \caption{The AP and precision of PointRCNN equipped with the LOP with different $B$ and different model structures.}
\scalebox{0.7}{
    \begin{tabular}{lcccc}
    \toprule
          % & \multicolumn{4}{c}{PointRCNN} \\
\cmidrule{2-5}          & AP    & \multicolumn{3}{c}{Precision} \\
\cmidrule(lr){2-2} \cmidrule(lr){3-5}          & Car   & Car   & Pedestrian & Cyclist \\
    \midrule
    w/o. defense  & 75.13\% & 75.04\% & 47.08\% & 56.87\% \\
    \midrule
    Ours(PointNet, B=0.2) & 75.96\% & 76.49\% & 50.70\% & 58.66\% \\
    Ours(PointNet, B=0.3) & 76.21\% & 77.15\% & \textbf{50.94\%} & 58.93\% \\
    Ours(PointNet, B=0.4) & \textbf{76.50\%} & 78.05\% & 50.49\% & 61.59\% \\
    Ours(PointNet, B=0.5) & 76.49\% & 79.29\% & 49.63\% & 61.92\% \\
    Ours(PointNet, B=0.6) & 76.37\% & \textbf{80.03\%} & 49.43\% & \textbf{63.87\%} \\
    \midrule
    Ours(DGCNN, B=0.2) & 76.44\% & 77.56\% & 51.04\% & 61.03\% \\
    Ours(DGCNN, B=0.3) & 76.66\% & 78.36\% & \textbf{51.31\%} & 61.72\% \\
    Ours(DGCNN, B=0.4) & 76.58\% & 79.44\% & 51.21\% & 63.74\% \\
    Ours(DGCNN, B=0.5) & 76.77\% & 80.75\% & 49.59\% & 65.14\% \\
    Ours(DGCNN, B=0.6) & \textbf{76.84\%} & \textbf{81.52\%} & 49.42\% & \textbf{67.24\%} \\
    \bottomrule
    \end{tabular}%
}
%   \vspace{-0.1in}
  \label{tab:ablation_normal}%
\end{table}%
%%%%%%%% END Ablation

\noindent\textbf{Results \& Analysis.}  
As we can see from Table \ref{tab:ablation_normal}, the choice of LOP's structure has limited influence on the performance of PointRCNN under the normal circumstance. The differences between their precision are at most $1.49\%$, $0.72\%$ and $3.37\%$ on cars, pedestrians and cyclists, and the differences between their AP are at most $0.48\%$ on cars.
In contrast, the value of $B$ greatly affects the performance of PointRCNN.
Normally, the higher value of $B$ is realted with better performance of PointRCNN with the LOP: the precision of PointRCNN on cars and on cyclists increase with a larger $B$, while the change of AP on cars is always less than $2\%$. 
% However, the LOP with a lower $B$ may bring PointRCNN better performance on detecting pedestrians, which is mainly because the larger precision of PointRCNN on pedestrians. As we can see from Fig.\ref{fig:ablation_atk}, although the LOP with the structure of DGCNN performs slightly better than the LOP with the structure of PointNet, the differences of the precision are always less than $3.48\%$, and the differences of the ASR are less than $4.00\%$ in most cases, a relatively small gap. Similarly, a higher $B$ always brings the better performance and the better defense effectiveness. For example, when $B=0.6$, the precision of PointRCNN with the LOP on cars increase \lyf{$10.79\%$ on} average, and the ASR of these appearing attacks decreases by $26.56\%$ on average compared with the results of $B=0.2$.

In fact, the point-wise PC model also performs well in other downstream tasks such as classification and semantic segmentation, which means the key features extracted by them is general enough to handle different CV tasks \cite{charles2017PN,charles2017PN++,yue2019DGCNN}. Thus, the LOP with different structures can both perform well in recognizing the components of real objects. However, in the pipeline of our proposed defense the value of $B$ directly determines whether a predicted object is preserved or eliminated. Therefore, the value of $B$ affects the performance of 3D object detectors equipped with the LOP.

% Besides, as a trade-off, the increase in $B$ always causes the degradation in recall in the normal circumstances and under attacks. However, the degradation is limited by $10\%$ on average. Based on our discussion in Section \ref{sec:Limitation}, we consider it as a reasonable trade-off because the slight decrease in recall has limited influence on the normal function of ADS in the real word. In summary, the effectiveness of our proposed defense is insensitive to different choices of the model structure of LOP, while the value of $B$ does play an important role on the contrary. Regardless of the performance of the 3D object detectors in the normal circumstances and the acceptable trading of recall, the LOP with a higher $B$ can always bring the 3D object detector larger improvement.
% \input{tex/app_realworld.tex}
\section{Mis-categorization Attack Experiments}
\label{appendix:miscategorization}

\noindent$\bullet$\textbf{ Experimental Settings.} To implement mis-categorization attack, we follow the idea of \textbf{Physical} attack: we collected the PCs about objects which was labeled as pedestrian in the training set of KITTI, and kept the PCs with less than $200$ points as the basic data of mis-categorization attack. Then, during each time of mis-categorization attacks, we randomly chose a PC from the basic data and injected it into the target sample, then we further used PGD to change the positions of some points in this PC in order to increased its confidence scores and its classification probability of vehicles. Table \ref{tab:EXP3} reports the ASR of this mis-categorization attack on 3 different object detectors, and the performance of these 3 object detectors with and without our LOP.

\begin{table*}[ht]
  \centering
  % \vspace{-0.2in}
  \caption{The ASR of mis-categorizan attacks and the performance of 3D object detectors under this attack.}
\scalebox{0.7}{
    \begin{tabular}{cccccccccc}
    \toprule
          & \multicolumn{3}{c}{PointPillars} & \multicolumn{3}{c}{PointRCNN} & \multicolumn{3}{c}{PV-RCNN} \\
    \cmidrule(lr){2-4} \cmidrule(lr){5-7} \cmidrule(lr){8-10}
          & ASR   & AP    & Precision & ASR   & AP    & Precision & ASR   & AP    & Precision \\
    \midrule
    w/o. defense & 4.42\% & 70.40\% & 73.78\% & 13.39\% & 64.36\% & 52.22\% & 12.86\% & 63.89\% & 28.51\% \\
    \midrule
    Ours (PointNet, B=0.5) & 3.67\% & 70.91\% & 78.80\% & 5.50\% & 72.09\% & 74.25\% & 6.44\% & 67.90\% & 61.24\% \\
    Ours (PointNet, B=0.6) & 3.36\% & 70.96\% & 79.50\% & 4.06\% & 72.85\% & 76.50\% & \textbf{5.81\%} & 68.01\% & 62.30\% \\
    Ours (DGCNN, B=0.5) & 3.39\% & \textbf{71.53\%} & 80.43\% & 4.67\% & 72.50\% & 75.88\% & 6.50\% & \textbf{68.47\%} & 63.96\% \\
    Ours (DGCNN, B=0.6) & \textbf{3.06\%} & 70.18\% & \textbf{81.30\%} & \textbf{3.44\%} & \textbf{73.18\%} & \textbf{77.83\%} & 5.89\% & 67.28\% & \textbf{65.01\%} \\
    \bottomrule
    \end{tabular}%
}
  \label{tab:EXP3}%
\end{table*}%

\noindent$\bullet$\textbf{ Results \& Analysis.} As we can see from Table \ref{tab:EXP3}, LOP reduces the ASR of the mis-categorization attack to almost half of its origin in most cases. For example, LOP reduces at least $58.92\%$ of the original ASR on PointRCNN and reduces at least $49.46\%$ of the original ASR on PV-RCNN. An exception is the PointPillars, which seems to be more resilient against mis-categorization attacks and thus the defense effect of LOP is not as clear as the other two cases. In addition, we also notice a similar phenomenon as discussed in Section  of our original manuscript, that LOP can slightly increase the performance of the 3D object detectors in some cases. For example, on PointPillars, the AP increase by $1.13\%$, while the precision increases $7.52\%$ when the detector is deployed with LOP. In summary, the experimental results validate that LOP is also effective against mis-categorization attacks, and incurs almost no overhead on the performance of object detectors.
\section{Fusion Experiments}
\label{appendix:fusion}
\noindent$\bullet$\textbf{ Experimental Settings.} According to \cite{Hallyburton2021SecurityAO}, we choose the fusion model EPNet, which can prevent appearing attacks in a certain degree, as the defense target. Based on the official implementation of EPNet, we implement the \textit{Physical} attack against EPNet, and evaluate its performance on KITTI with or without LOP. Table \ref{tab:EXP2-2} reports the corresponding results.
 
% Table generated by Excel2LaTeX from sheet 'Attack-Comp'
\begin{table}[ht]
  \centering
  % \vspace{-0.2in}
  \caption{The ASR of physical attack on EPNet and the performance of EPNet with and without LOP.}
\scalebox{0.7}{
    \begin{tabular}{cccc}
    \toprule
          & \makecell{Precision \\ (w/o. attack)} & \makecell{Precision \\ (w. attack)} & ASR \\
    \midrule
    None  & 46.45\% & 44.96\% & 44.39\% \\
    \midrule
    Ours(PointNet, B=0.4) & 61.82\% & 60.69\% & 0.53\% \\
    Ours(PointNet, B=0.5) & 64.19\% & 63.18\% & \textbf{0.00\%} \\
    Ours(PointNet, B=0.6) & \textbf{71.78\%} & \textbf{69.82\%} & \textbf{0.00\%} \\
    \midrule
    Ours (DGCNN, B=0.4) & 59.90\% & 57.28\% & 2.67\% \\
    Ours (DGCNN, B=0.5) & 62.37\% & 58.33\% & 0.53\% \\
    Ours (DGCNN, B=0.6) & 69.23\% & 67.05\% & 0.53\% \\
    \bottomrule
    \end{tabular}%
}
  \label{tab:EXP2-2}%
\end{table}%

\noindent$\bullet$\textbf{ Results \& Analysis.} As we can see from Table \ref{tab:EXP2-2}, without our LOP, the physical attacks can still forging objects in EPNet's detection and bring down the precision of it. However, the ASR of physical attack on EPNet can be reduced to less than $3\%$ with the support of our LOP, and the ASR will further be reduced to $0\%$ when we deploy PointNet-based LOP with $B=0.5,0.6$ on EPNet. Besides, the precision of EPNet with LOP also increased at least $12.32\%$ under attacks and at least $13.45\%$ in the normal circumstances. In summary, we believe LOP should not be viewed a competitor for multi-sensor fusion. Instead, LOP empirically improves the robustness of the fusion models and the 3D object detectors. As LOP has no interruption on the prediction on the image input and simply focus on eliminating malicious objects, LOP would not hurt the benefits of fusion models in self-driving systems.
\section{Details of The System-level Evaluation}
\label{appendix:system}

\begin{figure}[t]
    \centering
    \includegraphics[width=0.45\textwidth]{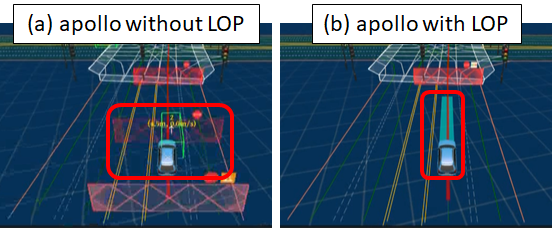}
    \caption{The Dreamview of Apollo without and with our LOP while the attack is aiming at interrupting the route planning.}
    \label{fig:sys_level_apollo}
\end{figure}

Specifically, to perform these experiments, we first select 5 different fake PCs which are located in front of the self-driving vehicle and can successfully forge the perception module of Apollo or those 3D object detectors for at least 1 frame in the previous experiments. 
% Meanwhile, we use Apollo to record the traces of the self-driving vehicle moving in 3 different maps. 
Then, we inject the selected fake PCs into 15 testing traces to conduct the appearing attack against the self-driving vehicle. 

We use the Dreamview to visualize the generated future routes of Apollo 6.0.0, which are shown as the green rectangles in front of the self-driving vehicle and represent the moving trajectories of the self-driving vehicle under the Apollo's control. Based on these routes and the planning module, we define that the self-driving vehicle is harsh braking when the planning module guide the self-driving vehicle decelerate to 0 km/h in less than 1 second.

As shown in the part (a) of Fig.\ref{fig:sys_level_apollo}, the generated future route is extended to the crossroads, which means the self-driving vehicle will move normally and stop at a red light under the instructions of the Apollo with our LOP. Therefore, we consider this situation as ``normal''. Meanwhile, as shown in the part (b) of Fig.\ref{fig:sys_level_apollo}, the generated future route disappears for a while, and the planning module guide the self-driving vehicle stop in a certain place, which means the self-driving vehicle will falsely brake in the middle of the road. Therefore, we consider this situation as a ``harsh braking''. We calculate the proportion of the ``harsh braking'' in the 15 poisoned traces as the harsh braking rate, and report it in Section \ref{sec:Experiment:results:real}.

% Specifically, we record the trace of the self-diving car and calculate the crash rate based on the planning route shown in the Dreamview. As shown in Fig.\ref{fig:sys_level_apollo}, the Apollo without LOP will be misled by the attacks and the planning module will falsely suggest to stop behind the forged car, which may further lead to the car crash between the self-driving car and the vehicles behind it. At the same time, the Apollo with LOP can identify the forged car and ignore it in the perception module, so its planning module can correctly generating the route during the whole trace and avoid the accidents.

% \input{tex/app_metrics.tex}
% \input{tex/app_method.tex}

%%%%%%%%%%%%%%%%%%%%%%%%%%%%%%%%%%%%%%%%%%%%%%%%%%%%%%%%%%%%%%%%%%%%%%%%%%%%%%%%
\end{document}